\documentclass{emulateapj}

\submitted{Accepted for publication in ApJ}

\shorttitle{Quasars in the COSMOS Field}
\shortauthors{Prescott et al.}

\def\ang{$\hbox{\AA$\!$}$}

\begin{document}

\title{Quasars in the COSMOS Field}

\author{M. K. M. Prescott\altaffilmark{1}, C. D. Impey\altaffilmark{1}, R. J. Cool\altaffilmark{1}, and N. Z. Scoville\altaffilmark{3}}

\altaffiltext{1}{Steward Observatory, 933 North Cherry Avenue, The University of Arizona, Tucson, AZ 85721; mprescott@as.arizona.edu, cimpey@as.arizona.edu, rcool@as.arizona.edu}
\altaffiltext{3}{Robinson Laboratory, California Institute of Technology, Pasadena, CA 91125; nzs@phobos.caltech.edu}

\begin{abstract}
We obtained medium-resolution spectra of 336 quasar candidates in the 
COSMOS HST/Treasury field using the MMT 6.5-meter telescope 
and the Hectospec multi-object spectrograph.  
Candidates were drawn from the Sloan Digital Sky Survey (SDSS) DR1 catalog using quasar flags set by 
the SDSS multi-color quasar target selection algorithm.  
In this paper we present our discovery spectra from 
1.39 square degrees (69.5\% of the COSMOS field) and a discussion of the 
selection method and yields.  
We confirmed 95 quasars, including at least 2 BALs; 80 of these are new quasars that do not appear in 
previous quasar confirmation follow-up studies.  
The candidates additionally included 184 compact emission-line galaxies, a 
third of which are likely Type 2 AGN, and 12 stars.  
The quasars span a range 
in magnitude of $18.3 < g < 22.5$ and a range in redshift of $0.2 < z < 2.3$.    
Our results are consistent with a lower limit quasar surface 
density from SDSS color selection of 102 per square degree down to $g = 22.5$ 
over the entire COSMOS field.  
This work is the first step toward the 
eventual goal of setting up a grid of quasar absorption line probes 
of the 2 square degree field, and of conducting a complete census of supermassive 
black holes in this well-studied survey region.  
The total quasar count at the conclusion 
of this study is 139, making COSMOS one of the most 
densely-sampled regions of sky where a grid of quasar sightlines can be used to probe 
the intervening volume.

\end{abstract}


\section{INTRODUCTION}

Quasar absorption-line studies of the intergalactic medium have uncovered a complex
``web'' of cosmic structure.  Optimum studies of the cosmic web use a grid of distant quasars as a set of probes 
of the intervening matter.  Absorbing material along the line-of-sight is identified in 
the quasar spectrum with an equivalent width and redshift and can be used to study the distribution, 
chemical enrichment, and clustering relative to luminous matter in the same volume.  
The Lyman-$\alpha$ forest absorption traces 
highly ionized hydrogen of low column density and low chemical enrichment \citep{mey95}.  
These absorbers are weakly clustered
along the line-of-sight and are more closely associated with voids and large
scale structures than they are with individual bright galaxies \citep{gro98, bah93, wey98, pen02}. 
Metal absorption lines such as the Mg~\textsc{ii}~$\lambda\lambda$2796, 2800 and C~\textsc{iv}~$\lambda\lambda$1548, 1550 
doublets offer a complementary approach to the study of large scale structure, 
as they trace massive galaxy halos via the cross-section to metal-enriched gas.  

Unlike the targets of luminous matter surveys, quasar absorbers can be selected in ways that 
are largely unaffected by the cosmological $(1+z)^{-4}$ surface brightness dimming, 
surface brightness and morphology selection effects, $k-$corrections, and Malmquist bias.  
Thus, the advantage of absorbers as probes of the intergalactic medium is 
that they can be detected with an efficiency that does not depend strongly 
on redshift, given a sufficiently bright background quasar.  On the other hand, 
faint quasar surveys are required to set up a dense enough grid of probes in a 
contiguous area to sample the full range of cosmic structures.  
Obtaining a complete picture of the large scale structure involves unifying these disparate 
studies of dim matter in the IGM and luminous matter in galaxies.

The Cosmic Evolution Survey (COSMOS) is the largest HST survey 
ever undertaken with the ACS instrument \citep{sco06}.
The field is 1.4$^{\circ}\times$1.4$^{\circ}$
square, aligned E-W, N-S, centered at $10^{\rm h}00^{\rm m}28\fs6, +02^{\circ}12\arcmin21\farcs0$ (J2000),
and accessible to all major observatories. 
Extensive multi-wavelength observations of the field are in progress, 
including deep radio (VLA), UV (GALEX), X-ray (XMM), and infrared (Spitzer) mapping, and a densely-sampled 
galaxy redshift survey using the VLT. 
A total of 45,000 redshifts will be obtained within the 2 square degree region.
The field size was chosen to span the largest known cosmic structures at $1 < z < 2$, and 
the volume encompassed by the galaxy redshift survey rivals that of the Sloan Digital Sky Survey (SDSS).

Building on the extensive spatial, wavelength, and redshift coverage 
contained in the COSMOS field, the eventual goal of this project
is to build a large grid of quasars to yield 3D ``tomography'' of the IGM 
via Lyman-$\alpha$ absorption as well as 3D measures of the relationship 
between galaxy halos selected by Mg~\textsc{ii} and C~\textsc{iv} absorption and the visible baryons in luminous 
galaxies.  Similar observations of individual quasar sightlines 
have been used to this effect to show evidence of strong overdensities \citep{hei89} and that 
C~\textsc{iv} and Mg~\textsc{ii} have correlation power on scales up to $\sim$140 $h^{-1}_{70}$ Mpc \citep{loh01}. 
Absorber studies with multiple sightlines have been used to map out large scale three-dimensional structures: 
\citet{din96} found a flattened structure at $z \sim 2$ that is $\sim$40 $h^{-1}_{70}$ Mpc by 
110 $h^{-1}_{70}$ Mpc while \citet{wil96} discovered a 20-50 $h^{-1}_{70}$ Mpc structure 
toward the South Galactic Pole.  Cohen et al. (1996, 1999) 
found that roughly half of the bright galaxies in deep galaxy redshift surveys are located in high-contrast structures 
separated by 70-430 $h^{-1}_{70}$ Mpc along the line-of-sight.  Multiple sightlines can also be 
used to measure three-dimensional 
absorber clustering \citep{sar87}; Lyman-$\alpha$ absorbers seem to be weakly clustered 
on scales of 30-40 $h^{-1}_{70}$ Mpc \citep{wil00, lis00}. 
The distribution of diffuse IGM baryons 
traced by Lyman-$\alpha$ and metal-line absorbers can be compared to the distribution of 
dark matter traced by weak lensing and the large scale structure defined 
by galaxy redshift surveys.  

This project must be approached in several stages.    
The first phase is the confirmation of quasars within the COSMOS field.  Color-selection methods and 
existing surveys provide the candidate population, and low-resolution 
spectra are sufficient for identification of the broad quasar emission 
features (e.g. Lyman-$\alpha$, C~\textsc{iv}, C~\textsc{iii}, and Mg~\textsc{ii}).  The second phase involves 
high resolution spectroscopic follow-up of confirmed quasars 
providing equivalent widths and identifications of absorbers along the line-of-sight 
toward the quasar.  In the third and final phase these absorber data, tracing the diffuse IGM and galaxy halos, 
will be merged with the COSMOS dark matter and galaxy redshift surveys in order to map out the 
3D distribution of dim and luminous matter.  There is now the prospect of using COS on the HST to map Lyman-$\alpha$ 
at $0 < z < 1.6$, below the atmospheric cut-off, and of comparing directly to galaxy redshifts measured with the VLT.

The goal of the current work is to begin setting 
up a grid of quasars across the COSMOS field using pre-existing quasar samples and selection methods.  
This paper presents the results of 
our first quasar confirmation run using a large sample of previously untargeted 
quasar candidates selected from the SDSS DR1 photometric catalog \citep{york00, abaz03}.  
More recent work within the extended COSMOS collaboration are yielding larger and 
fainter quasar samples \citep{imp06, tru06}
In Section 2 we discuss our sample selection, in Section 3 our 
observations and reduction procedures.  Section 4 and 5 present our 
results and analysis, and we review our conclusions in Section 6.
Discovery spectra are shown in Appendix A.

\section{SAMPLE SELECTION}

We derived our target sample from the SDSS DR1 photometric catalog.  
As part of the SDSS quasar target selection, objects contained in the SDSS 
photometric catalogs were assigned a number of 
flags related to their likelihood of being a quasar and whether or not they 
were to be targeted for follow-up SDSS spectroscopy.  These determinations were 
based on finding outliers from the stellar locus in 
4-color space and matching SDSS sources against the FIRST catalog of 
radio sources.  The relevant flags are 
described by \citet{rich02a} and references therein.  
\citet{van05} find empirically that the completeness of the SDSS selection 
algorithm is 94.9\% down to the limiting magnitude of $i = 19.1$, while that of the 
whole SDSS quasar survey is 89\% due to image defects, unidentifiable spectra, 
and extended sources.  
Objects selected as quasar candidates that lie outside the 
SDSS spectroscopic magnitude limits of $15 < i^* < 19.1$ for low-redshift 
and $15 < i^* < 20.2$ for high-redshift selection are flagged accordingly 
and left to follow-up studies such as the current work.  

Using the SDSS quasar flags, we compiled a list 
of quasar candidates, removing the 48 objects that had previously been targeted by SDSS (as of DR1), 
resulting in 1391 targets over the entire COSMOS field.  
We then made two cuts, keeping only those 
targets with $u$-band rms errors $< 0.3$ magnitudes and $g < 22.5$.  
Since photometric errors are highest 
in the $u$-band, and since ultraviolet excess is a key discriminant of a quasar, 
we made the $u$-band error cut in order to remove targets with unreliable photometry.  
The $g$ magnitude limit was chosen as the 
faintest magnitude bin in which the 
candidates were well-defined in a $u-g$ versus $g-r$ plot relative to typical stellar colors.  
After applying these cuts, we were left with 566 quasar candidate targets.  

Figure 1 shows a map of the 566 target positions within the COSMOS field; 
a plot of $g$-magnitude versus $u-g$ color for the final target sample is 
given in Figure 2.  The 43 quasars confirmed within the COSMOS field from SDSS DR1 and the additional 21 (of which 20 overlap our sample) 
confirmed up to and including the SDSS DR4 spectroscopic follow-up are shown in open circles \citep{adel06}.  
Compared to the candidate sample 
for the present study, they are confined to magnitudes brighter than $g = 20.5$ 
and slightly bluer $u-g$ colors.  Similarly, Figure 3 shows the SDSS $g-r$ versus $u-g$ 
plot for our target sample 
and the previously-confirmed SDSS quasars.  The two populations cover 
similar regions of color-color space, with our target sample extending to redder $g-r$ colors.  
From this plot it is clear our sample is not missing regions of color-color space 
with respect to the previously-known and brighter SDSS quasar sample.

Magnitudes quoted in this study are observed SDSS magnitudes, uncorrected for Galactic extinction.  
For reference, the mean Galactic extinction for this field is $A_{g}\sim0.07$ from Schlegel et al. (1998).
The typical photometric error for SDSS $g$ and $r$ magnitudes is $\sim$0.25.  

\section{OBSERVATIONS AND REDUCTIONS}

\subsection{Observational Details} 

Observations were carried out during the night of 2004 April 18 using the MMT 6.5-meter telescope and 
the Hectospec multi-object spectrograph \citep{fab04}.  Hectospec has 300 
optical fibers, each 1.5 arcseconds in diameter, which are positioned by two fiber robots to an accuracy of $\sim$0.2 arcseconds in $\sim$5 minutes. 
The resolution of the grating selected for this Hectospec set-up was 6~\ang.  The 
minimum fiber spacing is roughly 20 arcseconds, but the actual fiber constraints for a 
particular observational set-up are determined by the 
fiber positioning software.  The peak final throughput for Hectospec, including the telescope optics, is 21\% at 5000~\ang.  
The Hectospec detector array is composed of two 2048$\times$4608 pixel CCDs, 
with the gap positioned parallel to the dispersed spectrum.  
The grating used had 270 line mm$^{-1}$ and was blazed at 5200~\ang.  
The full spectral coverage was 5800~\ang, spanning from $\sim$3100-9000~\ang, at a dispersion of 1.21~\ang\,~pixel$^{-1}$.
Normal calibration frames were taken (dome flat, comparison arc lamp, 
and bias frames); no flux standards were necessary for this initial quasar confirmation phase.  

The targets were divided between two overlapping pointings 
as shown in Figure 1, which were observed for 60 and 80 minutes, respectively.  
Of the 566 SDSS quasar candidates in the COSMOS field, 388 were within the two pointings, 
and we obtained spectra of 336.  
The number of targets and sky fibers per pointing, pointing positions, 
individual integration times, and airmasses are given in Table~1.  As the field was setting during 
our observations, we used a longer integration time on 
the second pointing to partially compensate for the increase in airmass; the fractional quasar yields 
for the two pointings are similar: 0.27 for pointing I and 0.28 for pointing II.

Conditions during the observations were mostly clear, but image quality was affected by 
high winds.  Two problems with the data were noticed after the run $-$ 
an LED leak affecting the red end of the spectra ($\sim$8000-9000~\ang) 
and a problem with the atmospheric dispersion corrector (ADC) causing lowered counts, particularly in the blue spectral region 
($\sim$3000-4000~\ang).  Fortunately, these issues presented only minor problems and did not prevent us from achieving the goals 
of this phase of the project.

\subsection{Data Reduction} 
We reduced the Hectospec data with HSRED\footnote{http://mizar.as.arizona.edu/rcool/hsred/index.html}, an IDL
package developed for reduction of data from the Hectospec
and Hectochelle instruments on the MMT.  The code draws
heavily on the reduction pipeline used by SDSS.  The 300
fiber trace locations were determined using dome flat
spectra.  These dome flat observations were further used to
correct for the high frequency flat field variations in the
observations and the CCD fringing.  Sky subtraction was
performed using dedicated sky fibers distributed throughout
each field.  The bright sky lines in each spectrum were used
to refine the wavelength solution determined from a set of
HeNeAr comparison spectra obtained at the beginning of the night.

Object classifications and redshifts were determined using a
modified version of the SDSS redshift pipeline available
publicly in the IDLSPEC2D package.  Each spectrum was
compared to a grid of galaxy, quasar, and star spectra. A
$\chi^{2}$ minimization method was used to determine the best-fit
spectral classification and redshift.

Verification by eye confirmed all 
but a handful of cases in which low signal-to-noise, target faintness, or the spectral 
problems described in Section 3.1 caused the cross-correlation method to fail.  For 
these objects we measured the redshifts by hand using a flux-weighted mean 
and a simple Monte Carlo error estimation method.  All objects were assigned a 
classification of QUASAR, QUASAR1, QUASAR?, QUASAR BAL, GALAXY, GALAXY1, GALAXY?, 
STAR, STAR?, or UNKNOWN.  Table 2 gives a description of the observational criteria 
for each of these categories.  
Objects with secure classifications are all included in the 
final numbers for the overall categories of QUASAR, GALAXY, 
and STAR, while those with question marks and those in the UNKNOWN 
category remain without a secure identification and are not used in any subsequent statistical analysis.

In many cases the LED leak and ADC error had an almost purely cosmetic effect on the data. 
However, for targets with redshifts between 0.3 and 0.5, the Mg~\textsc{ii} emission line lies 
in the compromised blue end of the spectrum and H$\alpha$ is located in the compromised 
red end, leaving only H$\beta$ and the [O~\textsc{iii}] doublet lines for determining a redshift.  
In these cases, the combined loss of reliable spectral 
coverage meant that a few additional objects could not be confidently identified or have their 
redshifts measured. 

\section{RESULTS}

From these MMT observations we confirmed 95 quasars (including at least 2 BALs), 184 emission-line galaxies, and 
12 stars.  Of these, 14 quasars were previously confirmed by the 2dF survey and by more recent  
(post-DR1) SDSS spectroscopic follow-up (noted in Table 3); one additional quasar was confirmed by 2dF alone.  
This study therefore contributes 80 additional faint quasars in the COSMOS field.   
A total of 45 targets remain without a secure classification 
at the conclusion of this study, 
but only 7 objects were completely unidentifiable.  
Table 3 gives a list of confirmed objects along with redshift 
measurements where applicable.  All object spectra and classifications for confirmed targets are given in Figures 13.1-13.56 of Appendix A, 
presented in order of Right Ascension.  In all, this small field now has 139 confirmed or probable quasars, 
one of the highest concentrations of spectroscopically confirmed quasars in the sky.

The two panels of Figure 4 give the SDSS $g$-magnitude 
differential and cumulative histograms of the target sample, divided into four categories: 
the full sample for the 2 square degree COSMOS field, the sample contained in the area 
observed, the objects actually targeted, and the raw quasar yields from our run.  
In Figure 5 we plot our measured success rate per 1 magnitude bin; we use a linear fit 
to predict the number of quasars we would have confirmed if we had been able to cover the 
full SDSS quasar candidate sample.  We expect that 
if we had been able to target all candidates located within the two pointings, we would have 
recovered 110 total quasars, 15 of which would overlap the SDSS DR4 list.  Of the overlaps, 14 are included in the 
current study while 1 additional object was not observed as part of this work.  
We project that the total quasar count over the full field using 
the SDSS selection flags would have been roughly 159, 20 of which would overlap the SDSS DR4 list. 
Thus, in two pointings 
and a little over two hours of MMT/Hectospec observation time we were able to confirm nearly 
60\% of the quasars that had not been previously-targeted for spectroscopic follow-up (as of SDSS DR1) 
down to $g = 22.5$ over 2 square degrees.  

Combining our prediction of 110 quasars for the observed pointings with the 25
non-overlapping quasars confirmed by SDSS, 
we calculate a surface density of 97$\pm$8 quasars per square degree down to 
SDSS $g = 22.5$.  Over the entire COSMOS field, we combine our prediction of 159 quasars plus 
the 44 non-overlapping quasars confirmed by SDSS and calculate an expected surface density of 102$\pm$7 
per square degree. These surface densities and the projected quasar yield 
are lower limits, as some fraction of the unclassified objects may turn out to be quasars.  If all 
14 QUASAR? objects are in fact quasars, we predict 127 quasars over our two pointings (with 
15 SDSS overlaps) and a surface density in the observed portion of the field 
of 109$\pm$9 per square degree.  The projected quasar yield would be roughly 184 (with 20 SDSS overlaps) 
for the entire COSMOS field, resulting in a projected surface density estimate of 114$\pm$8 per square degree.  
Most likely only a fraction of the QUASAR? objects are in fact quasars, so these predictions are upper limits.  

Table 4 summarizes these results, while Figure 6 shows the combined differential and cumulative histograms, 
where we have included our newly-confirmed faint SDSS sample and previously-confirmed SDSS quasars.  
For comparison, \citet{rich05} present the 2dF-SDSS LRG and QSO Survey (2SLAQ) cumulative number counts 
to be 93.8 per square degree down to $g = 22$ (magnitude corrected for Galactic extinction) 
and for a redshift range of $0.3 < z < 2.2$, 
limits very similar to our survey.  For the same constraints, work by \citet{boy00} corresponds to 79 per square degree and 
work by \citet{cro04} yields 63 per square degree.  
A rough extrapolation of the cumulative quasar histogram of \citet{rich05} predicts that we should see 
between 90 and 100 per square degree down to $g\sim22.5$, consistent with our results.

Some fraction of the narrow-line objects are likely to be low luminosity Type 2 AGN.  
For confirmed emission-line galaxies in our sample at redshifts below 0.38, when the H$\alpha$ and NII lines remain within 
our observed wavelength range, we measure the flux in the [N~\textsc{ii}]~$\lambda$6584, H$\alpha$, [O~\textsc{iii}]~$\lambda$5007, and H$\beta$ emission lines.  
In Figure 7 we plot the flux ratios on a BPT diagram \citep{bal81}.  
Using the revised relation from \citet{kau03} for distinguishing 
star-forming galaxies and AGN and a signal-to-noise ratio (SNR) cut of 3, we find that roughly 30 out of 111 are in fact Type 2 objects.  
We define our SNR as the flux in the line divided by the rms scatter in the continuum near the line.  
This fraction of $\sim$0.27 translates into a prediction that $\sim$50 of our 184 confirmed galaxies are in fact Type 2 objects.  

\section{DISCUSSION}
\subsection{Optimizing Quasar Selection}
The SDSS quasar flags based on finding outliers from the stellar locus in 4-color space 
allowed us to select and confirm 95 quasars in the COSMOS field.  At brighter magnitudes ($g < 19$), the 
success rate was around 50\%, similar to our expectation from previous work using ultraviolet excess (UVX) quasar candidate 
selection and from the SDSS estimates.  \citet{mci04} used a slightly smaller sample of quasars 
to derive an overall UVX selection efficiency of 61\% down 
to $B \sim 22$.  \citet{rich02a} find an overall efficiency better than 65\% for the SDSS quasar target selection algorithm down 
to their limiting magnitude for spectroscopic follow-up of $i^{*} = 19.1$ for low-redshift and 
$i^{*} = 20.2$ for the high-redshift selection.
At dimmer magnitudes ($g > 19$), however, our fraction of confirmed quasars dropped steadily to below 15\%, in step with a 
rise in the primary contaminant population, emission-line galaxies.  The fraction of stars and unconfirmed 
objects also increased to dimmer magnitudes, particularly in the last magnitude bin $21.5 > g > 22.5$.  Figure 8 
demonstrates these trends graphically.  

Figure 9 shows the various sub-samples in color-color space.  As seen in Figure 3, the previously-confirmed 
SDSS quasars lie at bluer $g-r$ and $u-g$ colors.  Quasars discovered in the current work, along with the 
small number of stars, lie in the same region of color-color space as the previously-confirmed SDSS quasars, 
while the contaminant emission-line galaxy population corresponds to targets with substantially redder $g-r$ colors.  
This work suggests that when taken to fainter magnitudes the SDSS quasar flags select an increasingly 
large number of emission-line galaxies.  Why do so many emission-line galaxies make it into 
the SDSS-selected sample?  The primary reason is that the selection algorithm retains objects with the $g-r$ colors 
of a low-redshift quasar but redder than the typical quasar in our sample.  
In the upper left panel of Figure 10 we plot the $g-r$ histograms for the SDSS confirmed quasars and galaxies.  There 
is a clear offset in mean color between the two samples: 0.18 for the quasars and 0.67 for the galaxies.  

\citet{fan99} performed simulations of the colors and sizes of compact emission-line galaxies (CELGs) to predict 
how much they would contaminate the target pool produced by the SDSS quasar target selection algorithm.   
He found that the colors of CELGs are dominated by the continuum due 
to the large bandwidth of the SDSS filters, noting that an emission-line equivalent width (EW) of 100~\ang\, will alter 
the galaxy magnitude in a given filter by only 0.1 magnitudes.  Galaxies with extremely large 
EW indeed have peculiar colors, but they will show up in regions of color space far from 
quasars and normal stars.
Furthermore, he found that the CELG population will be redder on average than the quasars, due to 
the redder continuum, but in his simulations this effect is critically dependent on the choice of power-law distribution.  
\citet{hal96} suggested that a limit of $B-V < 0.6$ (or $g-r < 0.4$ using a \citet{fuk96} calibration) 
would eliminate half of the CELG population from the candidate pool.
\citet{fan99} concluded similarly that a cut of $g-r < 0.35$ would eliminate about 60\% of CELGs 
while retaining the largest number of quasars.  This cut is shown for reference in 
all four panels of Figure 10.

We can now ask the same question of our dataset:  what $g-r$ cut would have maximized quasar selection efficiency in our sample?  
The upper right panel of Figure 10 shows the number of quasars and galaxies retained in the sample for a particular $g-r$ 
cut; the cut retains all objects with $g-r$ less than the 
given value.  The number of quasars rises first for lower cut values but is quickly 
surpassed by the galaxy counts beyond values of $g-r = 0.6$.  
The lower left panel of Figure 10 gives the fraction of quasars, galaxies, and stars/unconfirmed objects 
for the same range of $g-r$ cuts.  There is an obvious and broad peak in the quasar fraction for 
cuts in the range $0.0 < g-r < 0.4$.  However, the ideal cut should maximize both the 
efficiency of the selection as well as the total number of quasars found.  
The lower right panel of Figure 10 plots the fraction of quasars and galaxies versus the number of quasars 
retained if a given $g-r$ cut is applied.  The point of maximum efficiency and maximum 
number of quasars occurs at a $g-r$ cut of $\sim 0.35$, in complete agreement with \citet{fan99} and 
\citet{hal96}.  This color cut reduces the number of emission-line galaxies 
from 184 to 20, i.e. it removes 89\% of the CELGs, an even more drastic improvement than 
achieved by \citet{hal96} and \citet{fan99}.  The cut also removes 16 quasars from our sample but 
increases the mean quasar fraction from 33\% to 57\%, a level in better agreement with previous UVX-selected 
studies done at brighter magnitudes.  
Even higher efficiency has been demonstrated by \citet{rich04} with the SDSS DR1 photometric data 
by applying a probability density analysis to training sets and using a nonparametric Bayesian classification.  
They achieve an efficiency of $\sim$95\% down to $g = 21$ while maintaining a completeness of 94.7\% to 
unresolved quasars brighter than $g \sim 19.5$.  At fainter magnitudes, however, \citet{rich04} expect 
reduced completeness. 

It should be noted that when working with state-of-the-art 
multi-object spectrographs over relatively moderate-sized fields 
optimizing the detection efficiency of a particular category of target 
is becoming an unnecessary or even unwanted precaution.  
The large number of fibers per square degree on the sky can compensate to some extent 
for a decrease in selection efficiency without costing additional observing time.  Furthermore, 
any simple cut will inevitably reduce the completeness of the survey; as mentioned above, 
the cut of $g-r < 0.35$ would have removed 16 (17\%) of the SDSS spectroscopically-confirmed quasars.  
For the current study, the goal is to find as many quasars as possible 
within one field in order to have a dense grid of absorption-line probes 
of the intervening matter; thus completeness is more important than efficiency.

\subsection{Emission-Line Galaxies}
The redshift distribution of our sample is shown in Figure 11.  The emission-line galaxies have a median 
redshift of $0.3$, while the median redshift of the quasars is $1.4$.  The redshift range of our sample of emission-line galaxies 
is $0.006 - 0.9$; the quasar sample spans redshifts from $0.2$ to nearly $2.3$.
The top panel of Figure 12 shows $g-r$ color versus redshift for the confirmed quasars and galaxies.  
While the quasars show a steady fall-off toward higher redshift, the emission-line galaxies 
peak in $g-r$ color around redshifts of $0.1 - 0.6$.  Again, it is this red, low-redshift population 
of CELGs that is the primary contaminant in this work.  
The lower panel of Figure 12 gives the SDSS $g$-band luminosity ($\nu L_{\nu}$) in ergs s$^{-1}$ versus redshift for both the quasar and galaxy samples.  
There is a separation between the quasars and galaxies, with quasars filling out the higher luminosity end and 
the galaxies residing at lower luminosities.  For reference, tracks of the redshift evolution 
of $L_{*}$ in the restframe $g$, $u$, 2800~\ang, and 1500~\ang\, bands are shown taken from work on the FORS Deep Field \citep{gab04}.  
The positions in redshift where these restframe bands pass through the observed $g$ band are shown as 
filled circles.  Thus, when observed in the g-band, an $L^{*}$ galaxy at $z\sim0.75$ would 
correspond in luminosity to the point labeled ``u''.  The galaxy population straddles $L^{*}$ at 
a redshift of $\sim0.4$.

\subsection{BAL Fraction}
The final topic of discussion is that of the fraction of broad-absorption line (BAL) quasars in our sample.  
Our spectra are not well-suited to a rigorous BAL selection 
algorithm involving continuum-fitting such as those used by \citet{tol02} and \citet{rei03} 
because of the lack of true flux calibration, the moderate to low SNR for the faint objects, 
and the cosmetic issues described in Section 3.  
However, from a visual inspection we would expect to be able to identify the most extreme cases, 
namely the LoBAL or LoBAL+HiBAL cases.  
LoBALs are quasars that show broad absorption troughs associated with low-ionization species such as Mg~\textsc{ii} as 
well as higher absorption species 
such as C~\textsc{iv} and C~\textsc{iii}.  Conversely, HiBAL quasars show broad absorption of high-ionization species only, 
while the LoBAL+HiBAL designates an intermediate 
category.  \citet{rei03} present composite spectra from SDSS of each of these classifications.  
From a visual inspection of the SDSS confirmed quasars in the COSMOS field, we find that 2 are likely LoBAL quasars, 
while none of the previous-confirmed quasars stand out as obvious LoBALs.
This corresponds to an observed LoBAL fraction of 1.7\% for our SDSS sample.

Due to the small sample we cannot 
derive the appropriate correction for selection effects; however, we note that \citet{rei03} find no 
sizable correction factor for BALs in the redshift range of our study ($z < 2.3$).  
We can therefore compare our rough LoBAL fraction with that of other studies; \citet{men01} find a 
corrected LoBAL fraction of 2.8\%$\pm$1.1\% or a more conservative estimate of 2.0\%$\pm$0.9\% using early SDSS data 
matched to the FIRST radio survey catalog.  
\citet{rei03} find a similar LoBAL fraction of 1.9$^{+0.5}_{-0.4}\%$ in the SDSS Early Data Release.  
Thus, our sample of quasars in the COSMOS field appears to be consistent with the LoBAL fractions quoted in 
previous SDSS studies.  Assuming a typical LoBAL to HiBAL distribution in the COSMOS 
field, we extrapolate that the results of a more thorough analysis of the overall BAL fraction would also be in agreement 
with these previous studies.

\section{CONCLUSIONS}

In this study we used the MMT and the Hectospec instrument to confirm a sample of 95 quasars in the 
2 square degree COSMOS field.  These quasar candidates were selected from the SDSS DR1 catalog using 
flags assigned by the SDSS multi-color quasar target selection algorithm.  Our findings are consistent with quasar surface densities of 102$\pm$7 per square degree over 
the entire COSMOS field down to $g = 22.5$.  From the current work, we find that the quasar fraction using 
the SDSS selection algorithm is 50\% for $g < 20$, 
falling to below 15\% at our magnitude limit of $g = 22.5$.  Of the 95 confirmed quasars, at least 2 are BALs.  
The primary contaminants to the sample are 184 emission-line galaxies with $g-r$ colors greater than 0.35 and 
redshifts between 0.2 and 0.6.  If efficiency 
is crucial in a color-selected survey, a cut retaining only objects with $g-r < 0.35$ can be applied, 
removing over 80\% of these compact emission-line galaxies and increasing the quasar yield to $\sim$60\% for magnitudes as faint as $g = 22.5$.

With a confirmed quasar population of 139, the COSMOS field is now one of the most densely sampled regions 
where a grid of sightlines can be used to identify absorbers as probes of a large contiguous volume.  
Ongoing work is extending the quasar-confirmation phase of this project down to $g\sim23.5$ 
\citep{tru06, imp06}.  Once a sample of quasar candidates exists for the entire field, 
the second phase of the project will involve using higher-resolution spectra of these quasar probes to detect 
absorbers along the line-of-sight.  These absorber data, which trace diffuse IGM and galaxy halos, will be 
combined with the COSMOS dark matter and galaxy redshift surveys in order to study the 
3D distribution of dim and luminous matter in the COSMOS field.

\acknowledgments

We are grateful to Cathy Petry and Xiaohui Fan for advice during the project 
and to Kris Eriksen, Andy Marble, and Daniel Eisenstein for observing assistance.
M. P. was supported by an NSF Graduate Research Fellowship.  
C. D. I. acknowledges support of this research through NASA award HST-GO-10092.04
to the University of Arizona.

Funding for the Sloan Digital Sky Survey (SDSS) has been provided by the Alfred P. Sloan Foundation, 
the Participating Institutions, the National Aeronautics and Space Administration, the National Science Foundation, 
the U.S. Department of Energy, the Japanese Monbukagakusho, and the Max Planck Society. The SDSS Web site is http://www.sdss.org/.

The SDSS is managed by the Astrophysical Research Consortium (ARC) for the Participating Institutions. The 
Participating Institutions are The University of Chicago, Fermilab, the Institute for Advanced Study, the 
Japan Participation Group, The Johns Hopkins University, Los Alamos National Laboratory, the Max-Planck-Institute 
for Astronomy (MPIA), the Max-Planck-Institute for Astrophysics (MPA), New Mexico State University, University of 
Pittsburgh, Princeton University, the United States Naval Observatory, and the University of Washington.

This research has made use of the NASA/IPAC Extragalactic Database (NED) which is operated by the Jet Propulsion Laboratory, 
California Institute of Technology, under contract with the National Aeronautics and Space Administration.


\clearpage

\begin{deluxetable}{lcc}
\tabletypesize{\scriptsize}
\tablecaption{Spectroscopic Observations}
\tablewidth{0pt}
\tablehead{
\colhead{Pointing} & \colhead{I} & \colhead{II}
}

\startdata
Right Ascension & 9:59:30.65 & 10:01:16.88 \\
Declination & +1:57:05.94 & +2:26:17.83 \\ 
Number of Targets & 153 & 183 \\
Number of Sky Fibers & 98 & 95 \\
Integration Time (s) & 3600 & 4800 \\
Mean Airmass & 1.16 & 1.28 \\
\enddata

\end{deluxetable}

\begin{deluxetable}{llcc}
\tabletypesize{\scriptsize}
\tablecaption{Spectroscopic Classification}
\tablewidth{0pt}
\tablehead{
\colhead{Classification} & \colhead{Description} & \colhead{Inclusion in Final Count} & \colhead{Number}
}

\startdata
QUASAR & Definite Quasar with 2 or more broad lines\tablenotemark{a} & QUASAR & 69 \\  
QUASAR1 & Definite Quasar with 1 broad line and additional piece of evidence\tablenotemark{b} & QUASAR & 24  \\  
QUASAR? & Uncertain, probably a Quasar & - & 14  \\  
QUASAR BAL & Broad Absorption Line Quasar & QUASAR & 2  \\
GALAXY & Definite Galaxy with 2 or more emission or absorption lines\tablenotemark{c} & GALAXY & 172  \\ 
GALAXY1 & Definite Galaxy with 1 narrow line and additional piece of evidence\tablenotemark{d} & GALAXY & 12  \\
GALAXY? & Uncertain, probably a Galaxy & - & 17  \\
STAR & Definite Star & STAR & 12  \\
STAR? & Uncertain, probably a Star & - & 7  \\
UNKNOWN & No classification was possible & - & 7 \\

\enddata

\tablenotetext{a}{Quasar broad emission line features, e.g. Ly$\alpha$, C~\textsc{iv}, C~\textsc{iii}, Mg~\textsc{ii}, H$\alpha$.}
\tablenotetext{b}{More uncertain but corroborating broad emission lines or spectral breaks outside the range 4000-8000 \ang.}
\tablenotetext{c}{The 4000 \ang\, break, CaII H \& K lines, galaxy emission lines, e.g. [OII], [OIII] doublet.}
\tablenotetext{d}{More uncertain but corroborating galaxy spectral features outside the range 4000-8000 \ang.}

\end{deluxetable}

\clearpage
\LongTables
\begin{deluxetable}{cccclc}
\tabletypesize{\scriptsize}
\tablecaption{Spectroscopically-Confirmed Targets in the COSMOS Field}
\tablewidth{0pt}
\tablehead{
\colhead{Name} & \colhead{RA (J2000)} & \colhead{Dec (J2000)} & \colhead{Classification} & \colhead{Redshift} & \colhead{Redshift Error$^{a}$} 
}

\startdata
SDSS J095739.25+020557.9 &  9:57:39.25 &  2:05:57.98 &  GALAXY &     0.322194 &     0.000014 \\ 
SDSS J095741.89+020126.2 &  9:57:41.89 &  2:01:26.25 &  GALAXY &     0.130360 &     0.000014 \\ 
SDSS J095743.44+015649.3 &  9:57:43.44 &  1:56:49.38 &  GALAXY &     0.575043 &     0.000026 \\ 
SDSS J095743.95+014631.3 &  9:57:43.95 &  1:46:31.36 &  GALAXY &     0.780036 &     0.000040 \\ 
SDSS J095746.21+015712.3 &  9:57:46.21 &  1:57:12.38 &  QUASAR &     0.904538 &     0.000425 \\ 
SDSS J095746.70+020711.7 &  9:57:46.70 &  2:07:11.71 &  QUASAR &     0.987245 &     0.000328 \\ 
SDSS J095754.77+015234.6 &  9:57:54.77 &  1:52:34.68 &  GALAXY &     0.194409 &     0.000013 \\ 
SDSS J095757.32+020119.0 &  9:57:57.32 &  2:01:19.05 &  GALAXY &     0.474040 &     0.000017 \\ 
SDSS J095758.11+020203.5 &  9:57:58.11 &  2:02:03.51 &  GALAXY &     0.209301 &     0.000009 \\ 
SDSS J095759.50+020435.9 &  9:57:59.50 &  2:04:35.90 &  QUASAR &     2.031420\tablenotemark{d} &     0.000191 \\ 
SDSS J095804.25+015526.5 &  9:58:04.25 &  1:55:26.58 &  GALAXY &     0.356245 &     0.000012 \\ 
SDSS J095804.43+020600.4 &  9:58:04.43 &  2:06:00.43 &  QUASAR &     1.843120 &     0.001279 \\ 
SDSS J095805.48+021657.9 &  9:58:05.48 &  2:16:57.93 &  GALAXY &     0.425338 &     0.000006 \\ 
SDSS J095806.99+014202.7 &  9:58:06.99 &  1:42:02.77 &  QUASAR &     1.710940 &     0.000600 \\ 
SDSS J095808.80+015620.9 &  9:58:08.80 &  1:56:20.94 &  GALAXY &     0.910153 &     0.000030 \\ 
SDSS J095809.92+021057.6 &  9:58:09.92 &  2:10:57.61 &  QUASAR &     0.836802 &     0.000420 \\ 
SDSS J095810.80+014218.7 &  9:58:10.80 &  1:42:18.72 &    STAR &     $\;\;\;\;\;\cdots$    &    $\cdots$    \\ 
SDSS J095811.08+013845.3 &  9:58:11.08 &  1:38:45.38 &  GALAXY &     0.393641 &     0.000030 \\ 
SDSS J095812.83+015525.0 &  9:58:12.83 &  1:55:25.03 &  GALAXY &     0.382592 &     0.000020 \\ 
SDSS J095813.31+013509.8 &  9:58:13.31 &  1:35:09.88 &  GALAXY &     0.124726 &     0.000019 \\ 
SDSS J095813.34+020536.0 &  9:58:13.34 &  2:05:36.02 &  QUASAR &     0.702752 &     0.000482 \\ 
SDSS J095815.29+014738.4 &  9:58:15.29 &  1:47:38.47 &  QUASAR &     2.212010 &     0.001236 \\ 
SDSS J095817.41+015922.7 &  9:58:17.41 &  1:59:22.77 &  GALAXY &     0.309371 &     0.000018 \\ 
SDSS J095817.53+021938.4 &  9:58:17.53 &  2:19:38.46 &  QUASAR &     0.730052 &     0.000242 \\ 
SDSS J095820.78+020213.4 &  9:58:20.78 &  2:02:13.41 &  QUASAR &     1.856180 &     0.000441 \\ 
SDSS J095825.55+013608.3 &  9:58:25.55 &  1:36:08.31 &  GALAXY &     0.322467 &     0.000022 \\ 
SDSS J095825.84+021002.2 &  9:58:25.84 &  2:10:02.24 &  GALAXY &     0.676363 &     0.000042 \\ 
SDSS J095827.76+014136.2 &  9:58:27.76 &  1:41:36.27 &  GALAXY &     0.006000 &     0.000006 \\ 
SDSS J095828.51+015645.8 &  9:58:28.51 &  1:56:45.85 &  GALAXY &     0.375036 &     0.000027 \\ 
SDSS J095829.04+014306.8 &  9:58:29.04 &  1:43:06.85 &  GALAXY &     0.733573 &     0.000066 \\ 
SDSS J095829.20+021542.7 &  9:58:29.20 &  2:15:42.73 &  QUASAR &     0.945722 &     0.000609 \\ 
SDSS J095829.80+021050.2 &  9:58:29.80 &  2:10:50.26 &  QUASAR &     1.188000 &     0.000316 \\ 
SDSS J095830.92+021127.0 &  9:58:30.92 &  2:11:27.06 &  GALAXY &     0.700667 &     0.000096 \\ 
SDSS J095830.96+013956.5 &  9:58:30.96 &  1:39:56.59 &  GALAXY &     0.671713 &     0.000039 \\ 
SDSS J095831.05+020153.3 &  9:58:31.05 &  2:01:53.32 &  GALAXY &     0.283694 &     0.000011 \\ 
SDSS J095837.65+015103.1 &  9:58:37.65 &  1:51:03.16 &  GALAXY &     0.310563 &     0.000028 \\ 
SDSS J095837.73+015144.5 &  9:58:37.73 &  1:51:44.56 &  GALAXY &     0.397999 &     0.000017 \\ 
SDSS J095841.79+015318.0 &  9:58:41.79 &  1:53:18.06 &  QUASAR &     1.783840 &     0.001197 \\ 
SDSS J095842.03+015857.7 &  9:58:42.03 &  1:58:57.72 &  QUASAR &     1.333690 &     0.000931 \\ 
SDSS J095842.27+013302.7 &  9:58:42.27 &  1:33:02.77 &  GALAXY &     0.633906 &     0.000070 \\ 
SDSS J095845.32+022309.5 &  9:58:45.32 &  2:23:09.56 &    STAR &     $\;\;\;\;\;\cdots$    &    $\cdots$    \\ 
SDSS J095847.66+020209.6 &  9:58:47.66 &  2:02:09.60 &  GALAXY &     0.340087 &     0.000031 \\ 
SDSS J095848.54+014922.4 &  9:58:48.54 &  1:49:22.40 &  GALAXY &     0.367708 &     0.000027 \\ 
SDSS J095851.50+015944.2 &  9:58:51.50 &  1:59:44.26 &  GALAXY &     0.202467 &     0.000030 \\ 
SDSS J095851.63+020020.6 &  9:58:51.63 &  2:00:20.66 &  GALAXY &     0.517379 &     0.000070 \\ 
SDSS J095852.45+021205.5 &  9:58:52.45 &  2:12:05.54 &  GALAXY &     0.379686 &     0.000034 \\ 
SDSS J095854.01+015601.6 &  9:58:54.01 &  1:56:01.68 &  GALAXY &     0.132268 &     0.000010 \\ 
SDSS J095854.72+013924.6 &  9:58:54.72 &  1:39:24.66 &  GALAXY &     0.130793 &     0.000032 \\ 
SDSS J095855.23+013509.3 &  9:58:55.23 &  1:35:09.38 &  GALAXY &     0.623221 &     0.000029 \\ 
SDSS J095856.27+014312.8 &  9:58:56.27 &  1:43:12.82 &  GALAXY &     0.350809 &     0.000019 \\ 
SDSS J095856.52+021255.2 &  9:58:56.52 &  2:12:55.22 &  GALAXY &     0.601531 &     0.000022 \\ 
SDSS J095857.35+020137.6 &  9:58:57.35 &  2:01:37.63 &  GALAXY &     0.241260 &     0.000011 \\ 
SDSS J095858.38+014705.3 &  9:58:58.38 &  1:47:05.38 &  GALAXY &     0.443308 &     0.000025 \\ 
SDSS J095858.43+015727.8 &  9:58:58.43 &  1:57:27.86 &  GALAXY &     0.187933 &     0.000042 \\ 
SDSS J095902.55+022511.4 &  9:59:02.55 &  2:25:11.42 &  QUASAR &     1.106050 &     0.000783 \\ 
SDSS J095902.76+021906.3 &  9:59:02.76 &  2:19:06.34 &  QUASAR &     0.345449\tablenotemark{d} &     0.000013 \\ 
SDSS J095903.23+022002.9 &  9:59:03.23 &  2:20:02.90 &  QUASAR &     1.138820 &     0.000595 \\ 
SDSS J095905.74+013827.3 &  9:59:05.74 &  1:38:27.31 &  GALAXY &     0.424361 &     0.000014 \\ 
SDSS J095906.46+013219.1 &  9:59:06.46 &  1:32:19.14 &  GALAXY &     0.364954 &     0.000015 \\ 
SDSS J095907.65+020820.6 &  9:59:07.65 &  2:08:20.65 &  QUASAR &     0.354505 &     0.000033 \\ 
SDSS J095912.18+020052.5 &  9:59:12.18 &  2:00:52.59 &  GALAXY &     0.399436 &     0.000016 \\ 
SDSS J095913.74+022222.8 &  9:59:13.74 &  2:22:22.80 &  GALAXY &     0.186780 &     0.000020 \\ 
SDSS J095915.06+014219.0 &  9:59:15.06 &  1:42:19.04 &  GALAXY &     0.445974 &     0.000013 \\ 
SDSS J095915.88+014430.2 &  9:59:15.88 &  1:44:30.26 &  GALAXY &     0.531081 &     0.000012 \\ 
SDSS J095917.26+015019.1 &  9:59:17.26 &  1:50:19.14 &  QUASAR &     1.342660 &     0.000246 \\ 
SDSS J095918.70+020951.4 &  9:59:18.70 &  2:09:51.48 &  QUASAR &     1.161600\tablenotemark{d} &     0.000345 \\ 
SDSS J095920.88+021431.0 &  9:59:20.88 &  2:14:31.05 &  GALAXY &     0.303490 &     0.000058 \\ 
SDSS J095920.89+020031.7 &  9:59:20.89 &  2:00:31.71 &  QUASAR &     1.483000 &     0.000999 \\ 
SDSS J095921.43+015847.0 &  9:59:21.43 &  1:58:47.06 &  GALAXY &     0.109154 &     0.000014 \\ 
SDSS J095921.85+013517.5 &  9:59:21.85 &  1:35:17.59 &  GALAXY &     0.186926 &     0.000028 \\ 
SDSS J095924.90+021444.4 &  9:59:24.90 &  2:14:44.41 &  GALAXY &     0.344956 &     0.000030 \\ 
SDSS J095925.76+013452.3 &  9:59:25.76 &  1:34:52.35 &    STAR &     $\;\;\;\;\;\cdots$    &    $\cdots$    \\ 
SDSS J095925.83+020231.0 &  9:59:25.83 &  2:02:31.09 &  GALAXY &     0.252403 &     0.000013 \\ 
SDSS J095927.40+022130.1 &  9:59:27.40 &  2:21:30.13 &  GALAXY &     0.133125 &     0.000018 \\ 
SDSS J095927.78+022224.3 &  9:59:27.78 &  2:22:24.34 &  GALAXY &     0.359376 &     0.000017 \\ 
SDSS J095928.33+021950.4 &  9:59:28.33 &  2:19:50.44 &  QUASAR &     1.478290 &     0.000538 \\ 
SDSS J095928.45+015703.4 &  9:59:28.45 &  1:57:03.42 &  GALAXY &     0.328650 &     0.000012 \\ 
SDSS J095928.45+015934.6 &  9:59:28.45 &  1:59:34.62 &  QUASAR &     1.165960 &     0.000671 \\ 
SDSS J095929.23+022034.2 &  9:59:29.23 &  2:20:34.22 &  QUASAR &     1.738060 &     0.001059 \\ 
SDSS J095929.41+015640.3 &  9:59:29.41 &  1:56:40.34 &  GALAXY &     0.561256 &     0.000023 \\ 
SDSS J095929.79+013531.6 &  9:59:29.79 &  1:35:31.63 &  GALAXY &     0.758662 &     0.000018 \\ 
SDSS J095932.14+022730.0 &  9:59:32.14 &  2:27:30.02 &  GALAXY &     0.332977 &     0.000018 \\ 
SDSS J095932.74+023841.6 &  9:59:32.74 &  2:38:41.64 &  GALAXY &     0.246618 &     0.000027 \\ 
SDSS J095934.00+021009.9 &  9:59:34.00 &  2:10:09.94 &  GALAXY &     0.186060 &     0.000008 \\ 
SDSS J095934.07+013707.2 &  9:59:34.07 &  1:37:07.28 &  GALAXY &     0.032031 &     0.000010 \\ 
SDSS J095934.68+021228.7 &  9:59:34.68 &  2:12:28.76 &  GALAXY &     0.345155 &     0.000017 \\ 
SDSS J095934.76+021551.9 &  9:59:34.76 &  2:15:51.94 &  GALAXY &     0.594848 &     0.000031 \\ 
SDSS J095934.89+021422.0 &  9:59:34.89 &  2:14:22.09 &  QUASAR &     1.733600 &     0.000768 \\ 
SDSS J095935.43+013059.7 &  9:59:35.43 &  1:30:59.79 &  QUASAR &     1.668750 &     0.000723 \\ 
SDSS J095938.07+022747.0 &  9:59:38.07 &  2:27:47.01 &  GALAXY &     0.426893 &     0.000025 \\ 
SDSS J095938.56+023316.7 &  9:59:38.56 &  2:33:16.77 &  QUASAR &     0.753602 &     0.000090 \\ 
SDSS J095938.91+023337.3 &  9:59:38.91 &  2:33:37.36 &  GALAXY &     0.412420 &     0.000072 \\ 
SDSS J095938.98+021201.1 &  9:59:38.98 &  2:12:01.18 &  QUASAR &     0.688950 &     0.000255 \\ 
SDSS J095939.22+022421.6 &  9:59:39.22 &  2:24:21.67 &  GALAXY &     0.224803 &     0.000035 \\ 
SDSS J095940.06+022306.6 &  9:59:40.06 &  2:23:06.64 &  QUASAR &     1.123300 &     0.000160 \\ 
SDSS J095940.26+015121.4 &  9:59:40.26 &  1:51:21.49 &  GALAXY &     0.250914 &     0.000006 \\ 
SDSS J095940.39+024145.3 &  9:59:40.39 &  2:41:45.38 &  GALAXY &     0.358390 &     0.000007 \\ 
SDSS J095940.74+021938.7 &  9:59:40.74 &  2:19:38.71 &  QUASAR &     1.459100 &     0.000378 \\ 
SDSS J095942.08+024103.1 &  9:59:42.08 &  2:41:03.12 &  QUASAR &        1.795\tablenotemark{c} &        0.002 \\ 
SDSS J095943.21+015216.6 &  9:59:43.21 &  1:52:16.60 &  GALAXY &     0.229967 &     0.000034 \\ 
SDSS J095944.10+024429.9 &  9:59:44.10 &  2:44:29.97 &  GALAXY &     0.464817 &     0.000013 \\ 
SDSS J095949.40+020140.9 &  9:59:49.40 &  2:01:40.98 &  QUASAR &     1.753520\tablenotemark{d} &     0.000451 \\ 
SDSS J095953.55+022738.9 &  9:59:53.55 &  2:27:38.98 &  GALAXY &     0.082599 &     0.000004 \\ 
SDSS J095954.78+013206.4 &  9:59:54.78 &  1:32:06.43 &  QUASAR &     0.481540 &     0.000062 \\ 
SDSS J095957.82+013208.8 &  9:59:57.82 &  1:32:08.88 &  GALAXY &     0.434306 &     0.000035 \\ 
SDSS J095958.15+014005.5 &  9:59:58.15 &  1:40:05.59 &  GALAXY &     0.091670 &     0.000004 \\ 
SDSS J095958.53+021805.1 &  9:59:58.53 &  2:18:05.18 &  QUASAR &     1.788850 &     0.000958 \\ 
SDSS J095958.54+024159.2 &  9:59:58.54 &  2:41:59.24 &  GALAXY &     0.728646 &     0.000024 \\ 
SDSS J095959.11+014941.5 &  9:59:59.11 &  1:49:41.59 &  GALAXY &     0.359542 &     0.000043 \\ 
SDSS J100000.17+015606.8 & 10:00:00.17 &  1:56:06.86 &  GALAXY &     0.371913 &     0.000023 \\ 
SDSS J100004.16+015146.4 & 10:00:04.16 &  1:51:46.40 &  GALAXY &     0.266362 &     0.000018 \\ 
SDSS J100005.59+024508.6 & 10:00:05.59 &  2:45:08.60 &  QUASAR &     0.500300 &     0.000115 \\ 
SDSS J100005.64+021607.8 & 10:00:05.64 &  2:16:07.89 &  GALAXY &     0.236488 &     0.000009 \\ 
SDSS J100006.80+022245.2 & 10:00:06.80 &  2:22:45.26 &  GALAXY &     0.011757 &     0.000005 \\ 
SDSS J100006.85+020139.5 & 10:00:06.85 &  2:01:39.54 &  GALAXY &     0.309486 &     0.000034 \\ 
SDSS J100006.97+020207.7 & 10:00:06.97 &  2:02:07.72 &  GALAXY &     0.314517 &     0.000012 \\ 
SDSS J100009.84+024226.9 & 10:00:09.84 &  2:42:26.92 &  GALAXY &     0.600963 &     0.000009 \\ 
SDSS J100009.96+014804.5 & 10:00:09.96 &  1:48:04.50 &  GALAXY &     0.267040 &     0.000025 \\ 
SDSS J100010.19+023744.9 & 10:00:10.19 &  2:37:44.94 &  QUASAR &        1.560\tablenotemark{b} &        0.002 \\ 
SDSS J100011.68+023158.0 & 10:00:11.68 &  2:31:58.08 &  GALAXY &     0.332432 &     0.000018 \\ 
SDSS J100011.84+022623.1 & 10:00:11.84 &  2:26:23.17 &  GALAXY &     0.440452 &     0.000007 \\ 
SDSS J100012.08+014439.8 & 10:00:12.08 &  1:44:39.87 &  QUASAR &     1.148090 &     0.000800 \\ 
SDSS J100012.91+023522.8 & 10:00:12.91 &  2:35:22.81 &  QUASAR &     0.701555\tablenotemark{d} &     0.000076 \\ 
SDSS J100013.70+013034.5 & 10:00:13.70 &  1:30:34.59 &  QUASAR &     0.852805 &     0.000240 \\ 
SDSS J100014.09+022838.5 & 10:00:14.09 &  2:28:38.56 &  QUASAR &     1.253000 &     0.000634 \\ 
SDSS J100014.46+013328.3 & 10:00:14.46 &  1:33:28.33 &  GALAXY &     0.344993 &     0.000017 \\ 
SDSS J100014.79+015426.1 & 10:00:14.79 &  1:54:26.10 &  GALAXY &     0.670886 &     0.000043 \\ 
SDSS J100016.25+022450.8 & 10:00:16.25 &  2:24:50.83 &  GALAXY &     0.350322 &     0.000014 \\ 
SDSS J100020.35+024228.9 & 10:00:20.35 &  2:42:28.90 &  GALAXY &     0.250147 &     0.000026 \\ 
SDSS J100021.74+015009.4 & 10:00:21.74 &  1:50:09.42 &  GALAXY &     0.218412 &     0.000022 \\ 
SDSS J100022.45+023018.0 & 10:00:22.45 &  2:30:18.03 &  GALAXY &     0.268133 &     0.000022 \\ 
SDSS J100023.31+023712.3 & 10:00:23.31 &  2:37:12.39 &  GALAXY &     0.291524 &     0.000022 \\ 
SDSS J100024.48+020619.6 & 10:00:24.48 &  2:06:19.62 &  QUASAR &     2.288520 &     0.000576 \\ 
SDSS J100024.64+023149.0 & 10:00:24.64 &  2:31:49.04 &  QUASAR &     1.311200\tablenotemark{d} &     0.000179 \\ 
SDSS J100024.71+025039.0 & 10:00:24.71 &  2:50:39.01 &  GALAXY &     0.696753 &     0.000051 \\ 
SDSS J100025.06+024128.4 & 10:00:25.06 &  2:41:28.46 &  QUASAR &     1.881820 &     0.000839 \\ 
SDSS J100025.25+015852.0 & 10:00:25.25 &  1:58:52.06 &  QUASAR &     0.372559\tablenotemark{d} &     0.000021 \\ 
SDSS J100025.30+024818.6 & 10:00:25.30 &  2:48:18.68 &  GALAXY &     0.331828 &     0.000016 \\ 
SDSS J100025.70+022202.8 & 10:00:25.70 &  2:22:02.85 &  GALAXY &     0.340500 &     0.000012 \\ 
SDSS J100026.56+014538.9 & 10:00:26.56 &  1:45:38.98 &  GALAXY &     0.569300 &     0.000234 \\ 
SDSS J100027.75+015703.9 & 10:00:27.75 &  1:57:03.99 &  GALAXY &     0.264465 &     0.000007 \\ 
SDSS J100029.07+022850.0 & 10:00:29.07 &  2:28:50.05 &  GALAXY &     0.688393 &     0.000050 \\ 
SDSS J100030.47+023735.5 & 10:00:30.47 &  2:37:35.54 &  QUASAR &     1.839460 &     0.000614 \\ 
SDSS J100031.61+014757.7 & 10:00:31.61 &  1:47:57.73 &  QUASAR &     1.679100 &     0.001215 \\ 
SDSS J100033.19+015038.4 & 10:00:33.19 &  1:50:38.47 &  GALAXY &     0.571064 &     0.000078 \\ 
SDSS J100033.39+015237.0 & 10:00:33.39 &  1:52:37.05 &  QUASAR &     0.832222 &     0.000273 \\ 
SDSS J100033.49+013811.4 & 10:00:33.49 &  1:38:11.40 &  QUASAR &     0.520990 &     0.000207 \\ 
SDSS J100033.79+024354.7 & 10:00:33.79 &  2:43:54.76 &  QUASAR &     1.310500 &     0.000667 \\ 
SDSS J100034.60+014649.9 & 10:00:34.60 &  1:46:49.98 &  GALAXY &     0.184455 &     0.000031 \\ 
SDSS J100034.93+020235.0 & 10:00:34.93 &  2:02:35.01 &  QUASAR &     1.179840 &     0.000828 \\ 
SDSS J100035.71+020113.4 & 10:00:35.71 &  2:01:13.44 &  GALAXY &     0.265439 &     0.000010 \\ 
SDSS J100035.86+022653.5 & 10:00:35.86 &  2:26:53.55 &  GALAXY &     0.338454 &     0.000009 \\ 
SDSS J100038.01+020822.3 & 10:00:38.01 &  2:08:22.38 &  QUASAR &        1.825\tablenotemark{c} &        0.002 \\ 
SDSS J100040.15+024751.5 & 10:00:40.15 &  2:47:51.50 &  QUASAR &     1.041570 &     0.000271 \\ 
SDSS J100040.70+024934.9 & 10:00:40.70 &  2:49:34.93 &  GALAXY &     0.416288 &     0.000031 \\ 
SDSS J100041.42+021331.8 & 10:00:41.42 &  2:13:31.87 &  GALAXY &     0.325041 &     0.000011 \\ 
SDSS J100042.09+022533.9 & 10:00:42.09 &  2:25:33.96 &  GALAXY &     0.311771 &     0.000009 \\ 
SDSS J100043.13+020637.2 & 10:00:43.13 &  2:06:37.22 &  QUASAR &     0.360599\tablenotemark{d} &     0.000039 \\ 
SDSS J100043.40+024054.4 & 10:00:43.40 &  2:40:54.44 &  GALAXY &     0.305230 &     0.000019 \\ 
SDSS J100043.64+014009.1 & 10:00:43.64 &  1:40:09.19 &  QUASAR &     2.029420 &     0.000349 \\ 
SDSS J100044.00+014942.6 & 10:00:44.00 &  1:49:42.63 &  GALAXY &     0.527891 &     0.000020 \\ 
SDSS J100045.03+020512.5 & 10:00:45.03 &  2:05:12.55 &    STAR &     $\;\;\;\;\;\cdots$    &    $\cdots$    \\ 
SDSS J100045.65+020123.2 & 10:00:45.65 &  2:01:23.26 &  GALAXY &     0.092532 &     0.000004 \\ 
SDSS J100046.73+020404.4 & 10:00:46.73 &  2:04:04.44 &  QUASAR &     0.553653 &     0.000166 \\ 
SDSS J100046.94+020015.8 & 10:00:46.94 &  2:00:15.87 &  QUASAR &     1.920230\tablenotemark{d} &     0.000707 \\ 
SDSS J100047.75+020756.9 & 10:00:47.75 &  2:07:56.96 &  QUASAR &     2.160590 &     0.001104 \\ 
SDSS J100049.81+015706.6 & 10:00:49.81 &  1:57:06.62 &  GALAXY &     0.361678 &     0.000029 \\ 
SDSS J100050.72+024925.1 & 10:00:50.72 &  2:49:25.10 &    STAR &     $\;\;\;\;\;\cdots$    &    $\cdots$    \\ 
SDSS J100050.95+024753.5 & 10:00:50.95 &  2:47:53.59 &  GALAXY &     0.429036 &     0.000018 \\ 
SDSS J100051.10+024248.4 & 10:00:51.10 &  2:42:48.49 &  GALAXY &     0.186305 &     0.000023 \\ 
SDSS J100051.51+021215.3 & 10:00:51.51 &  2:12:15.30 &  QUASAR &     1.850060 &     0.000298 \\ 
SDSS J100051.93+015919.2 & 10:00:51.93 &  1:59:19.28 &  QUASAR &     2.239610 &     0.001088 \\ 
SDSS J100053.86+021220.6 & 10:00:53.86 &  2:12:20.62 &  GALAXY &     0.121853 &     0.000008 \\ 
SDSS J100054.84+022623.4 & 10:00:54.84 &  2:26:23.46 &  GALAXY &     0.464736 &     0.000017 \\ 
SDSS J100055.63+022150.4 & 10:00:55.63 &  2:21:50.40 &  QUASAR &     1.933090 &     0.000419 \\ 
SDSS J100056.24+021636.7 & 10:00:56.24 &  2:16:36.73 &  GALAXY &     0.123186 &     0.000013 \\ 
SDSS J100056.59+025129.5 & 10:00:56.59 &  2:51:29.55 &  GALAXY &     0.510724 &     0.000021 \\ 
SDSS J100058.33+015208.5 & 10:00:58.33 &  1:52:08.58 &  QUASAR &     2.024160 &     0.001954 \\ 
SDSS J100058.46+015436.9 & 10:00:58.46 &  1:54:36.97 &  GALAXY &     0.373979 &     0.000032 \\ 
SDSS J100058.70+022556.2 & 10:00:58.70 &  2:25:56.20 &  QUASAR &     0.695149 &     0.000384 \\ 
SDSS J100059.15+020647.4 & 10:00:59.15 &  2:06:47.48 &    STAR &     $\;\;\;\;\;\cdots$    &    $\cdots$    \\ 
SDSS J100059.49+021535.4 & 10:00:59.49 &  2:15:35.49 &    STAR &     $\;\;\;\;\;\cdots$    &    $\cdots$    \\ 
SDSS J100100.31+024413.7 & 10:01:00.31 &  2:44:13.74 &  QUASAR &     2.172130 &     0.000510 \\ 
SDSS J100100.74+014952.8 & 10:01:00.74 &  1:49:52.89 &  GALAXY &     0.528999 &     0.000048 \\ 
SDSS J100101.07+024115.7 & 10:01:01.07 &  2:41:15.79 &  GALAXY &     0.346603 &     0.000024 \\ 
SDSS J100104.07+023442.6 & 10:01:04.07 &  2:34:42.63 &    STAR &     $\;\;\;\;\;\cdots$    &    $\cdots$    \\ 
SDSS J100104.57+022627.2 & 10:01:04.57 &  2:26:27.20 &  GALAXY &     0.347905 &     0.000038 \\ 
SDSS J100104.67+025251.8 & 10:01:04.67 &  2:52:51.85 &  GALAXY &     0.452569 &     0.000028 \\ 
SDSS J100106.92+021929.2 & 10:01:06.92 &  2:19:29.20 &  GALAXY &     0.338016 &     0.000017 \\ 
SDSS J100109.82+020536.4 & 10:01:09.82 &  2:05:36.49 &  GALAXY &     0.282967 &     0.000013 \\ 
SDSS J100110.72+022049.1 & 10:01:10.72 &  2:20:49.12 &  GALAXY &     0.248073 &     0.000005 \\ 
SDSS J100110.75+020652.7 & 10:01:10.75 &  2:06:52.70 &  GALAXY &     0.247533 &     0.000032 \\ 
SDSS J100110.77+025320.2 & 10:01:10.77 &  2:53:20.25 &  GALAXY &        0.103\tablenotemark{b} &        0.002 \\ 
SDSS J100111.88+014535.6 & 10:01:11.88 &  1:45:35.64 &  GALAXY &     0.339773 &     0.000051 \\ 
SDSS J100111.92+023024.7 & 10:01:11.92 &  2:30:24.76 &  QUASAR &     1.494800 &     0.000513 \\ 
SDSS J100112.24+015842.2 & 10:01:12.24 &  1:58:42.24 &  GALAXY &     0.388559 &     0.000025 \\ 
SDSS J100112.62+020939.8 & 10:01:12.62 &  2:09:39.85 &  QUASAR &     1.817170 &     0.000492 \\ 
SDSS J100114.76+015748.4 & 10:01:14.76 &  1:57:48.49 &  GALAXY &     0.360802 &     0.000033 \\ 
SDSS J100115.61+015052.4 & 10:01:15.61 &  1:50:52.44 &  GALAXY &     0.567047 &     0.000060 \\ 
SDSS J100116.29+023607.5 & 10:01:16.29 &  2:36:07.52 &  QUASAR &     0.959422 &     0.000644 \\ 
SDSS J100117.13+025255.7 & 10:01:17.13 &  2:52:55.77 &  GALAXY &     0.694750 &     0.000073 \\ 
SDSS J100118.52+015542.8 & 10:01:18.52 &  1:55:42.81 &  QUASAR &     0.528661 &     0.000045 \\ 
SDSS J100118.57+022958.0 & 10:01:18.57 &  2:29:58.05 &  GALAXY &     0.217207 &     0.000013 \\ 
SDSS J100118.62+021431.1 & 10:01:18.62 &  2:14:31.12 &    STAR &     $\;\;\;\;\;\cdots$    &    $\cdots$    \\ 
SDSS J100119.32+024559.6 & 10:01:19.32 &  2:45:59.68 &  GALAXY &     0.073011 &     0.000007 \\ 
SDSS J100120.18+023047.7 & 10:01:20.18 &  2:30:47.70 &  GALAXY &     0.176016 &     0.000009 \\ 
SDSS J100120.24+022339.3 & 10:01:20.24 &  2:23:39.37 &  GALAXY &     0.247721 &     0.000014 \\ 
SDSS J100120.25+020341.2 & 10:01:20.25 &  2:03:41.25 &  QUASAR &     0.905615 &     0.000289 \\ 
SDSS J100121.24+014738.6 & 10:01:21.24 &  1:47:38.61 &  GALAXY &     0.339145 &     0.000026 \\ 
SDSS J100123.12+020310.5 & 10:01:23.12 &  2:03:10.58 &  GALAXY &     0.460211 &     0.000025 \\ 
SDSS J100123.17+023930.9 & 10:01:23.17 &  2:39:30.96 &  GALAXY &     0.219494 &     0.000023 \\ 
SDSS J100128.51+024114.5 & 10:01:28.51 &  2:41:14.53 &  GALAXY &     0.419649 &     0.000087 \\ 
SDSS J100129.28+021116.2 & 10:01:29.28 &  2:11:16.29 &  GALAXY &     0.569147 &     0.000085 \\ 
SDSS J100129.66+020643.3 & 10:01:29.66 &  2:06:43.30 &  QUASAR &     1.916270 &     0.001279 \\ 
SDSS J100132.21+020147.1 & 10:01:32.21 &  2:01:47.10 &  GALAXY &     0.501262 &     0.000026 \\ 
SDSS J100132.81+015759.7 & 10:01:32.81 &  1:57:59.79 &  QUASAR &     1.529660 &     0.000598 \\ 
SDSS J100133.78+023303.0 & 10:01:33.78 &  2:33:03.06 &  GALAXY &     0.112673 &     0.000014 \\ 
SDSS J100135.10+023223.1 & 10:01:35.10 &  2:32:23.10 &  GALAXY &     0.113566 &     0.000034 \\ 
SDSS J100135.45+025406.0 & 10:01:35.45 &  2:54:06.08 &  QUASAR &     1.642370 &     0.001126 \\ 
SDSS J100135.58+024341.5 & 10:01:35.58 &  2:43:41.52 &  GALAXY &     0.916410 &     0.000055 \\ 
SDSS J100136.50+025303.5 & 10:01:36.50 &  2:53:03.58 &  QUASAR &     2.112940\tablenotemark{d} &     0.000625 \\ 
SDSS J100136.52+024222.9 & 10:01:36.52 &  2:42:22.96 &  GALAXY &     0.487841 &     0.000028 \\ 
SDSS J100137.59+022155.6 & 10:01:37.59 &  2:21:55.69 &  QUASAR &     2.039950 &     0.001609 \\ 
SDSS J100140.37+024332.3 & 10:01:40.37 &  2:43:32.37 &  GALAXY &     0.223378 &     0.000028 \\ 
SDSS J100140.68+015931.1 & 10:01:40.68 &  1:59:31.12 &  GALAXY &     0.266793 &     0.000032 \\ 
SDSS J100141.53+023459.7 & 10:01:41.53 &  2:34:59.73 &  GALAXY &     0.822187 &     0.000045 \\ 
SDSS J100141.84+020200.2 & 10:01:41.84 &  2:02:00.20 &  GALAXY &     0.684676 &     0.000037 \\ 
SDSS J100141.91+022714.5 & 10:01:41.91 &  2:27:14.50 &  GALAXY &     0.174935 &     0.000042 \\ 
SDSS J100142.50+023923.0 & 10:01:42.50 &  2:39:23.04 &  GALAXY &     0.550259 &     0.000034 \\ 
SDSS J100144.11+023346.2 & 10:01:44.11 &  2:33:46.22 &  GALAXY &     0.403245 &     0.000017 \\ 
SDSS J100144.48+021626.9 & 10:01:44.48 &  2:16:26.97 &  GALAXY &     0.474159 &     0.000013 \\ 
SDSS J100147.89+021447.1 & 10:01:47.89 &  2:14:47.14 &  QUASAR &     0.881799 &     0.000288 \\ 
SDSS J100148.44+020841.1 & 10:01:48.44 &  2:08:41.17 &  QUASAR &     1.300850 &     0.000722 \\ 
SDSS J100148.99+024821.7 & 10:01:48.99 &  2:48:21.70 &  QUASAR &     1.612920 &     0.000665 \\ 
SDSS J100149.66+021512.6 & 10:01:49.66 &  2:15:12.60 &  GALAXY &     0.424917 &     0.000024 \\ 
SDSS J100150.08+024150.6 & 10:01:50.08 &  2:41:50.60 &  GALAXY &     0.195230 &     0.000010 \\ 
SDSS J100150.54+023119.5 & 10:01:50.54 &  2:31:19.56 &  GALAXY &     0.329404 &     0.000039 \\ 
SDSS J100151.11+020032.4 & 10:01:51.11 &  2:00:32.43 &  QUASAR &     0.966629 &     0.000297 \\ 
SDSS J100151.49+022939.0 & 10:01:51.49 &  2:29:39.04 &  GALAXY &     0.187481 &     0.000055 \\ 
SDSS J100152.90+022913.0 & 10:01:52.90 &  2:29:13.05 &  GALAXY &     0.434961 &     0.000015 \\ 
SDSS J100153.28+022436.8 & 10:01:53.28 &  2:24:36.82 &  QUASAR &     0.666530 &     0.000477 \\ 
SDSS J100155.75+020153.4 & 10:01:55.75 &  2:01:53.47 &  GALAXY &     0.320069 &     0.000023 \\ 
SDSS J100156.79+025312.1 & 10:01:56.79 &  2:53:12.19 &  GALAXY &     0.624114 &     0.000187 \\ 
SDSS J100157.42+024932.0 & 10:01:57.42 &  2:49:32.01 &  GALAXY &     0.522543 &     0.000018 \\ 
SDSS J100157.78+024631.6 & 10:01:57.78 &  2:46:31.65 &  QUASAR &     0.035760 &     0.000226 \\ 
SDSS J100158.53+023839.3 & 10:01:58.53 &  2:38:39.37 &  GALAXY &     0.195412 &     0.000038 \\ 
SDSS J100158.95+022445.2 & 10:01:58.95 &  2:24:45.21 &  QUASAR &     1.370600 &     0.001008 \\ 
SDSS J100159.21+025211.3 & 10:01:59.21 &  2:52:11.35 &  GALAXY &     0.744574 &     0.000021 \\ 
SDSS J100159.42+023935.6 & 10:01:59.42 &  2:39:35.64 &  QUASAR &     0.851119 &     0.000481 \\ 
SDSS J100159.78+022641.7 & 10:01:59.78 &  2:26:41.71 &  QUASAR &     2.032930 &     0.000318 \\ 
SDSS J100201.36+023629.7 & 10:02:01.36 &  2:36:29.77 &  GALAXY &     0.213058 &     0.000017 \\ 
SDSS J100201.51+020329.4 & 10:02:01.51 &  2:03:29.48 &  QUASAR &     2.060500\tablenotemark{d} &     0.000159 \\ 
SDSS J100202.45+023054.7 & 10:02:02.45 &  2:30:54.79 &  GALAXY &     0.219060 &     0.000032 \\ 
SDSS J100202.59+024524.2 & 10:02:02.59 &  2:45:24.26 &  GALAXY &     0.675794 &     0.000019 \\ 
SDSS J100208.57+020328.0 & 10:02:08.57 &  2:03:28.08 &  QUASAR &     1.199730 &     0.000286 \\ 
SDSS J100214.03+022559.7 & 10:02:14.03 &  2:25:59.70 &  GALAXY &     0.501816 &     0.000037 \\ 
SDSS J100216.54+024214.6 & 10:02:16.54 &  2:42:14.65 &  GALAXY &     0.351628 &     0.000024 \\ 
SDSS J100216.58+022018.4 & 10:02:16.58 &  2:20:18.49 &  GALAXY &     0.394484 &     0.000023 \\ 
SDSS J100217.42+022959.7 & 10:02:17.42 &  2:29:59.74 &  QUASAR &     1.106370 &     0.000489 \\ 
SDSS J100217.45+023545.9 & 10:02:17.45 &  2:35:45.96 &  GALAXY &     0.220081 &     0.000027 \\ 
SDSS J100217.81+024321.7 & 10:02:17.81 &  2:43:21.79 &  GALAXY &     0.331244 &     0.000037 \\ 
SDSS J100219.01+022242.4 & 10:02:19.01 &  2:22:42.42 &  GALAXY &     0.222120 &     0.000013 \\ 
SDSS J100219.42+025049.5 & 10:02:19.42 &  2:50:49.52 &  QUASAR &     1.450950 &     0.000218 \\ 
SDSS J100219.47+021315.9 & 10:02:19.47 &  2:13:15.96 &  QUASAR &     2.027330 &     0.000467 \\ 
SDSS J100221.45+022938.7 & 10:02:21.45 &  2:29:38.76 &  GALAXY &     0.259461 &     0.000011 \\ 
SDSS J100222.24+024225.4 & 10:02:22.24 &  2:42:25.48 &  GALAXY &     0.323868 &     0.000030 \\ 
SDSS J100225.74+024028.0 & 10:02:25.74 &  2:40:28.09 &  GALAXY &     0.094013 &     0.000005 \\ 
SDSS J100229.15+020931.8 & 10:02:29.15 &  2:09:31.82 &  QUASAR &     1.517140 &     0.000587 \\ 
SDSS J100231.43+023942.7 & 10:02:31.43 &  2:39:42.76 &  GALAXY &     0.505580 &     0.000011 \\ 
SDSS J100232.13+023537.3 & 10:02:32.13 &  2:35:37.32 &  QUASAR &     0.657166\tablenotemark{d} &     0.000088 \\ 
SDSS J100232.29+022939.5 & 10:02:32.29 &  2:29:39.58 &    STAR &     $\;\;\;\;\;\cdots$    &    $\cdots$    \\ 
SDSS J100234.85+024253.1 & 10:02:34.85 &  2:42:53.17 &  QUASAR &     0.195840\tablenotemark{d} &     0.000086 \\ 
SDSS J100241.83+022525.3 & 10:02:41.83 &  2:25:25.35 &  GALAXY &     0.359499 &     0.000018 \\ 
SDSS J100241.88+023314.0 & 10:02:41.88 &  2:33:14.04 &  GALAXY &     0.186241 &     0.000011 \\ 
SDSS J100244.65+021152.8 & 10:02:44.65 &  2:11:52.83 &    STAR &     $\;\;\;\;\;\cdots$    &    $\cdots$    \\ 
SDSS J100246.33+021917.0 & 10:02:46.33 &  2:19:17.04 &  GALAXY &     0.178724 &     0.000060 \\ 
SDSS J100246.50+024549.4 & 10:02:46.50 &  2:45:49.46 &  GALAXY &     0.881450 &     0.000033 \\ 
SDSS J100249.33+023746.5 & 10:02:49.33 &  2:37:46.52 &  QUASAR &     2.131780 &     0.000603 \\ 
SDSS J100249.92+021732.3 & 10:02:49.92 &  2:17:32.31 &  QUASAR &     1.092280 &     0.001263 \\ 
SDSS J100249.95+022330.1 & 10:02:49.95 &  2:23:30.19 &  GALAXY &     0.485346 &     0.000017 \\ 
SDSS J100250.20+023849.0 & 10:02:50.20 &  2:38:49.05 &  GALAXY &     0.094168 &     0.000012 \\ 
SDSS J100251.63+022905.3 & 10:02:51.63 &  2:29:05.38 &  QUASAR &     2.006750\tablenotemark{d} &     0.000215 \\ 
SDSS J100254.93+023515.6 & 10:02:54.93 &  2:35:15.61 &    STAR &     $\;\;\;\;\;\cdots$    &    $\cdots$    \\ 
SDSS J100255.03+020942.1 & 10:02:55.03 &  2:09:42.12 &  GALAXY &     0.325148 &     0.000020 \\ 
SDSS J100255.67+023025.4 & 10:02:55.67 &  2:30:25.41 &  GALAXY &     0.268544 &     0.000007 \\ 
SDSS J100255.69+023616.0 & 10:02:55.69 &  2:36:16.02 &  GALAXY &     0.375908 &     0.000015 \\ 
SDSS J100258.23+024141.3 & 10:02:58.23 &  2:41:41.38 &  GALAXY &     0.199750 &     0.000024 \\ 
SDSS J100304.02+022445.5 & 10:03:04.02 &  2:24:45.54 &  GALAXY &     0.371347 &     0.000015 \\ 
SDSS J100309.21+022038.3 & 10:03:09.21 &  2:20:38.32 &  QUASAR &     1.967030\tablenotemark{d} &     0.000762 \\ 
SDSS J100311.60+022803.0 & 10:03:11.60 &  2:28:03.00 &  GALAXY &     0.366929 &     0.000033 \\ 
SDSS J100315.99+022615.0 & 10:03:15.99 &  2:26:15.00 &  QUASAR &     2.031070 &     0.000367 \\ 

\enddata
\tablenotetext{a}{Redshift errors are formal errors returned by $\chi^2$ template-fitting procedure, and may be 
underestimates in the case of systematic shifts according to line species \citep{rich02b, mci99, tyt92}.}
\tablenotetext{b}{Redshift determined using a flux-weighted mean line centroid.  Redshift errors are $\Delta$z $\sim$ 0.002 or less.}
\tablenotetext{c}{BAL quasar with uncertain redshift.}
\tablenotetext{d}{Independently confirmed by the 2dF quasar survey or subsequent SDSS spectroscopic follow-up.}

\end{deluxetable}

\begin{deluxetable}{lcccc}
\tabletypesize{\scriptsize}
\tablecaption{Quasar Surface Density Projections}
\tablewidth{0pt}
\tablehead{
\colhead{} & \colhead{Observed Field}  & \colhead{COSMOS Field} 
}
\startdata
Number of Quasars (this study) & 110-127 &  159-184  \\ 
Number of Additional SDSS Quasars & 25 &  44 \\ 
Total Area (square degrees) & 1.39 & 2.00 \\
Surface Density at $g < 22.5$ (per square degree) & 97-109 & 102-114  \\
\enddata

\end{deluxetable}

\clearpage

\begin{figure}
\plotone{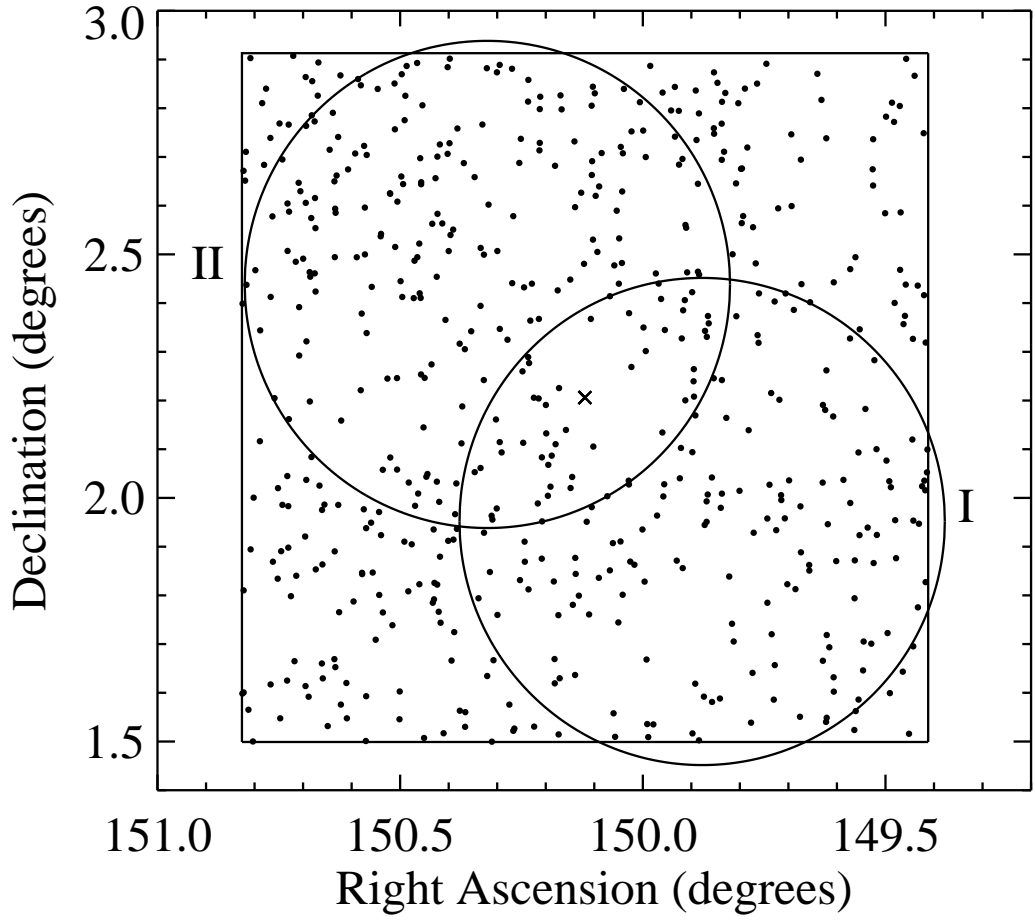}
\caption{Map of the spatial distribution of 566 SDSS quasar candidates within the 
2$\sq^{\circ}$ COSMOS field, overlaid with the 
two 1$^{\circ}$-diameter Hectospec pointings used during the observations.  
The COSMOS field is the inscribed square, with its center indicated as a cross at 
$10^{\rm h}00^{\rm m}28\fs6, +02^{\circ}12\arcmin21\farcs0$ (J2000).}
\end{figure}

\begin{figure}
\plotone{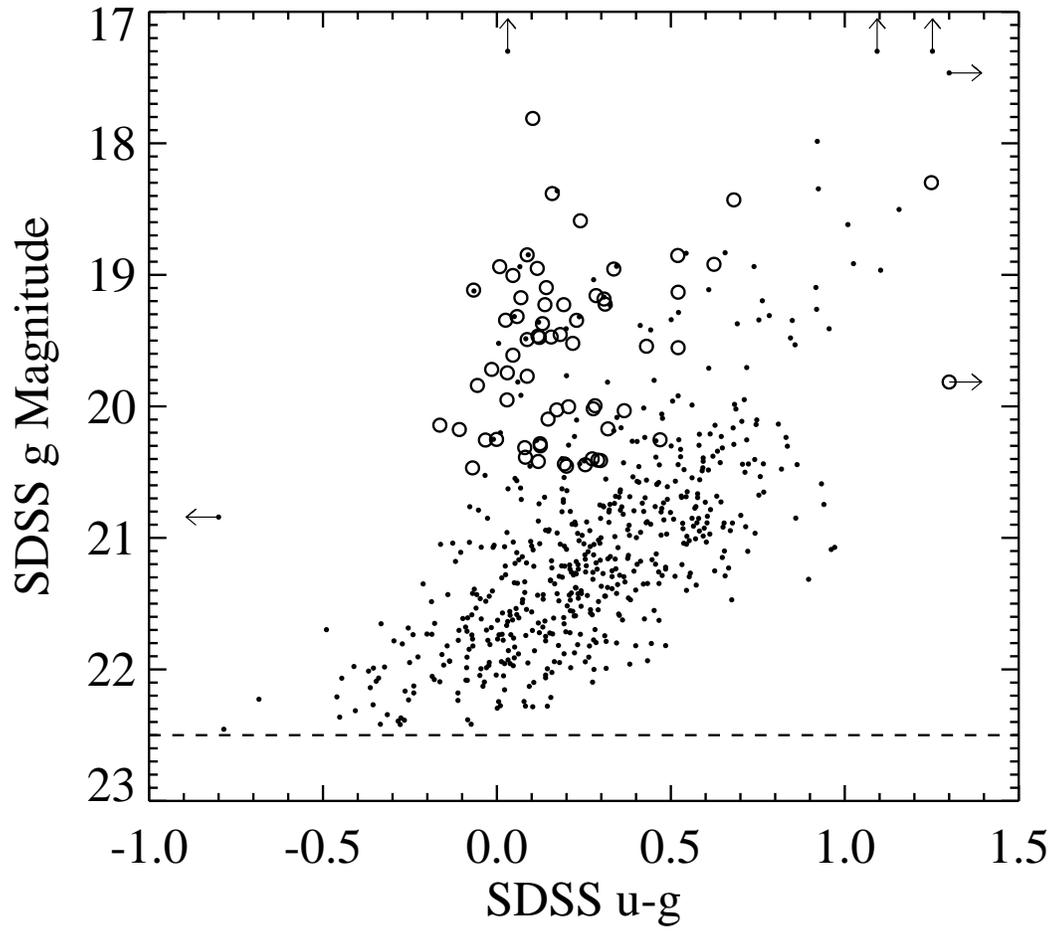}
\caption{Color-magnitude plot of the SDSS photometrically-selected targets (566 filled circles).  
The 64 quasars confirmed up to and including SDSS DR4 are shown as open circles.  Improvements in photometry between SDSS 
DR1 and DR4 accounts for the slight 
offsets visible in a few cases.  The dotted line indicates the $g < 22.5$ cut applied.}  
\end{figure}

\begin{figure}
\plotone{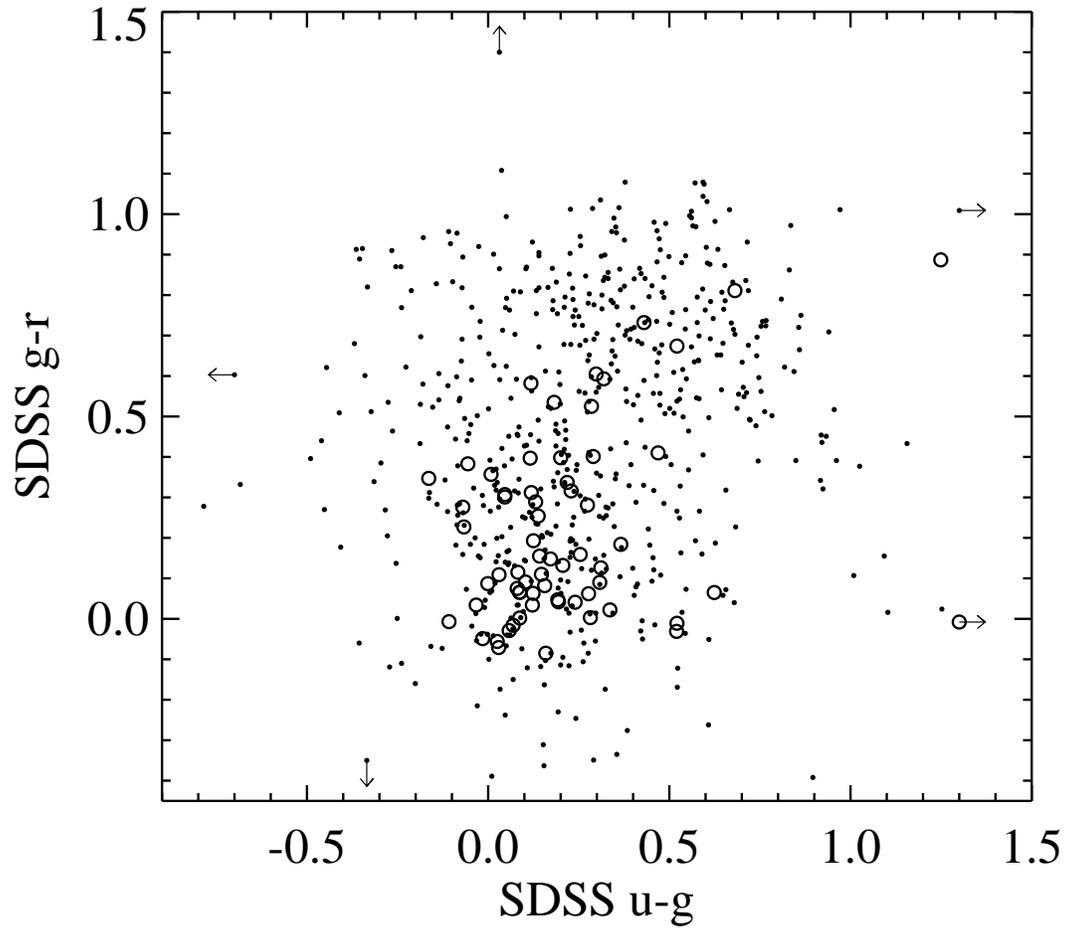}
\caption{Color-color plot of SDSS colors for quasar candidates in 
this study (566 filled circles) and previously confirmed SDSS quasars (64 open
circles).  Improvements in photometry between SDSS DR1 and DR4 account for 
the slight offsets that appear in a few cases.}  
\end{figure}

\begin{figure}
\plotone{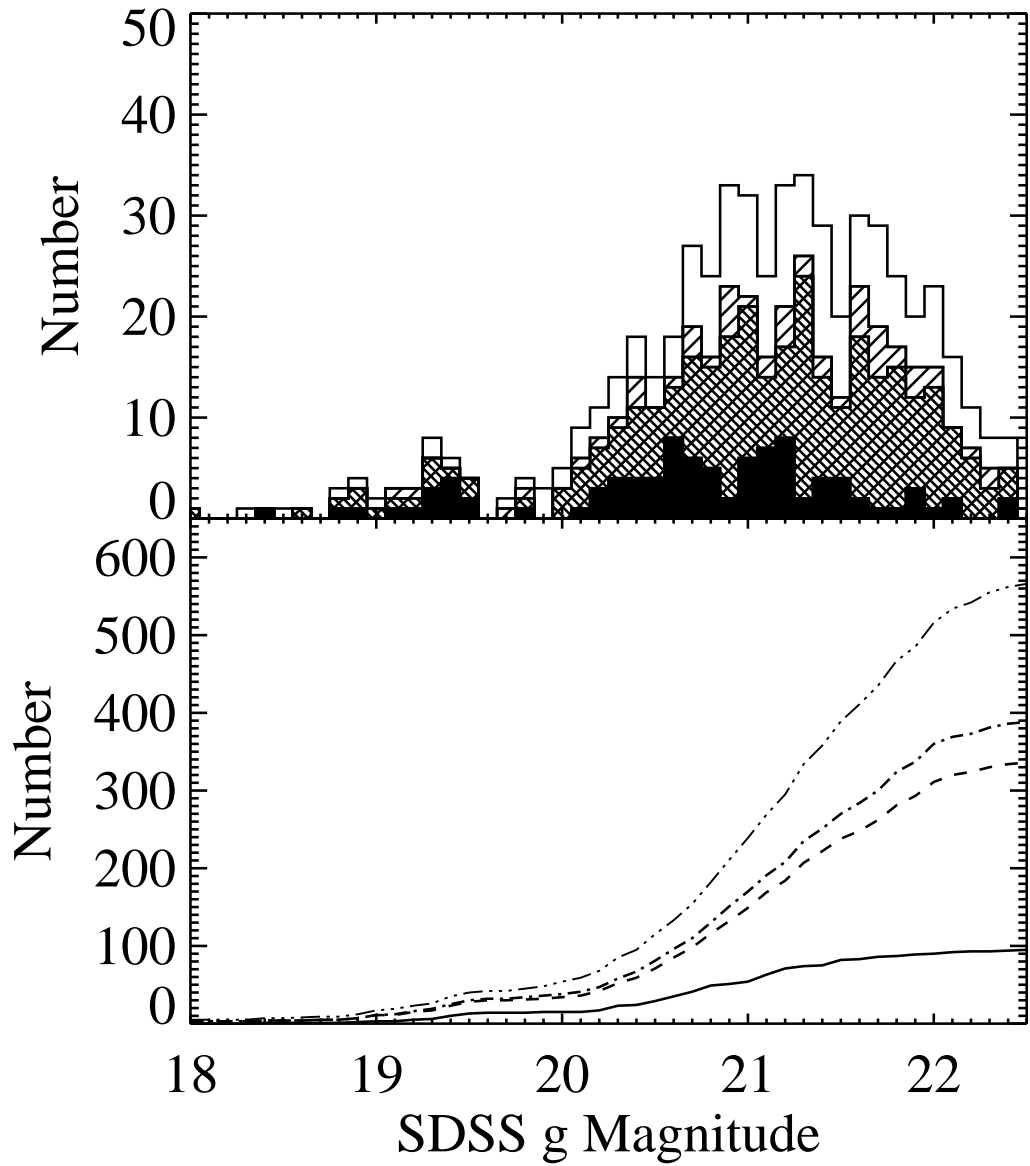}
\caption{(Top Panel) Histogram of SDSS quasar yield, in 0.1 magnitude bins.  
The empty layer contains all SDSS targets in the 2$\sq^{\circ}$ COSMOS field, 
the wide-hatch layer all SDSS targets within the pointing, the 
narrow-hatch layer all SDSS targets observed in this study, and the solid layer 
all SDSS targets confirmed in this study to be quasars.  The plot excludes five bright 
candidate outliers; three of the five were targeted, 
but none were confirmed to be quasars.  
(Bottom Panel) The same layers plotted as a cumulative histogram.}  
\end{figure}

\begin{figure}
\plotone{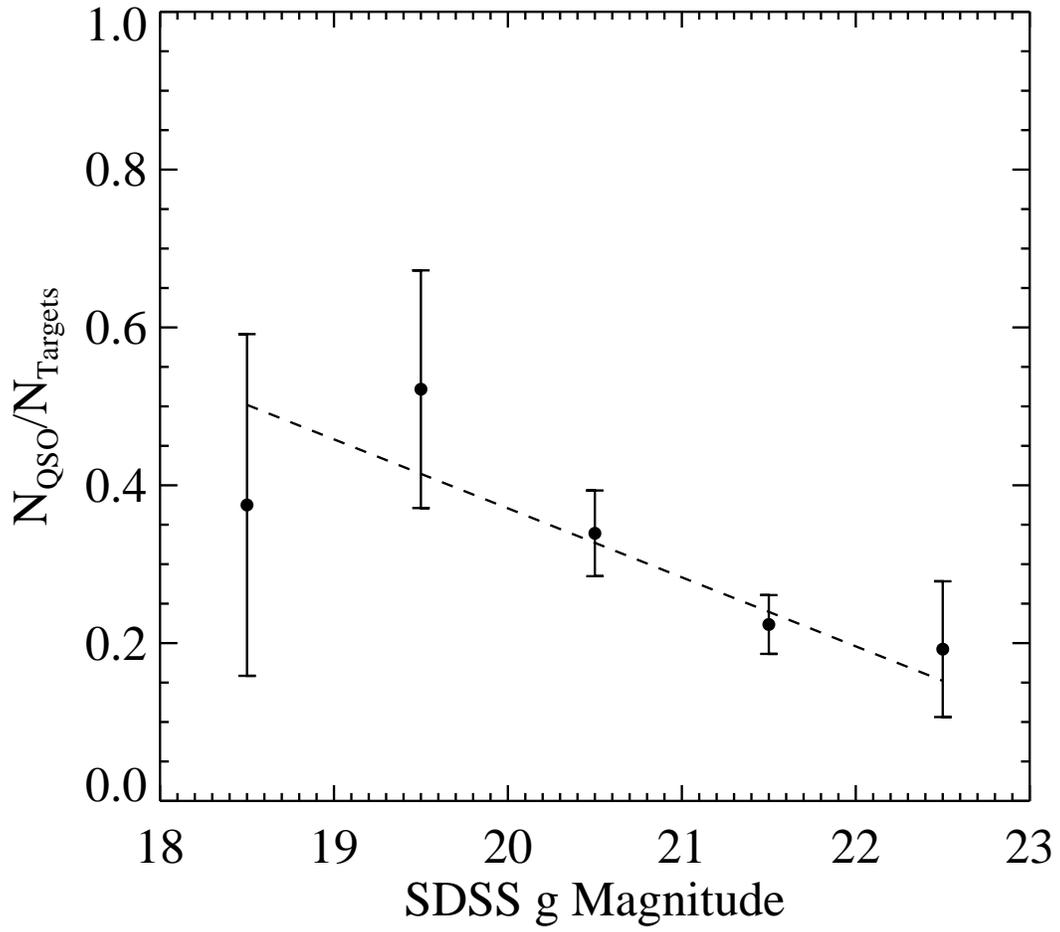}
\caption{Success rate of confirming quasars per 1.0 magnitude bin and the linear fit used to 
predict yields from the full target sample.}  
\end{figure}

\begin{figure}
\plotone{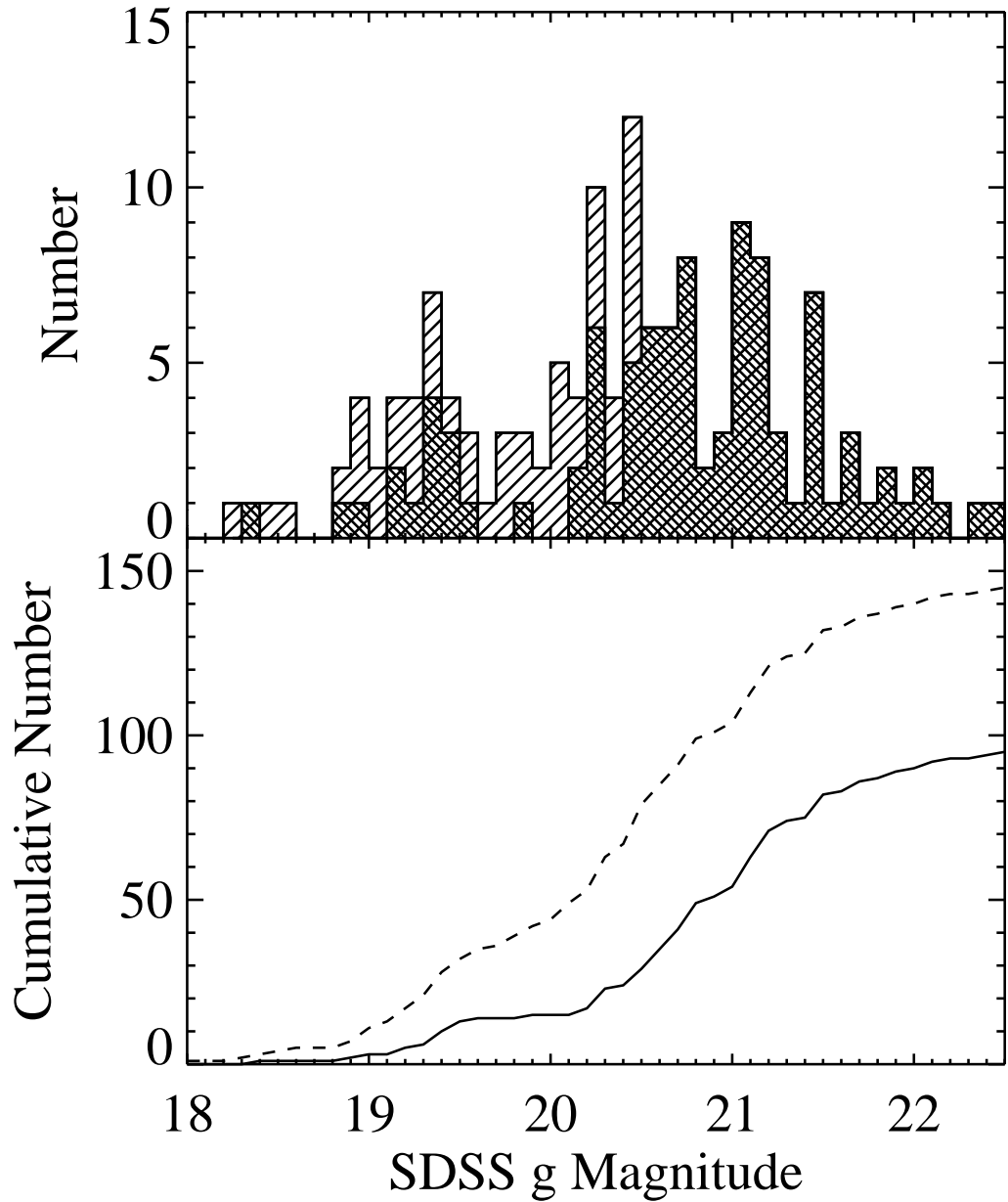}
\caption{(Top Panel)  Histogram of confirmed quasars in the COSMOS field.  
The narrow-hatch histogram indicate quasars confirmed only in this study.  
The wide-hatched histogram is a sum of all quasars confirmed in this field, both 
previously as part of SDSS spectroscopic follow-up and in the current work, taking into account sample overlaps.  
(Bottom Panel)  Cumulative histogram of confirmed quasars within the 2$\sq^{\circ}$ COSMOS field: quasars from 
this study alone are shown in the solid line while the dashed line includes 
confirmed quasars from both SDSS spectroscopic follow-up and the current work.}  
\end{figure}

\begin{figure}
\plotone{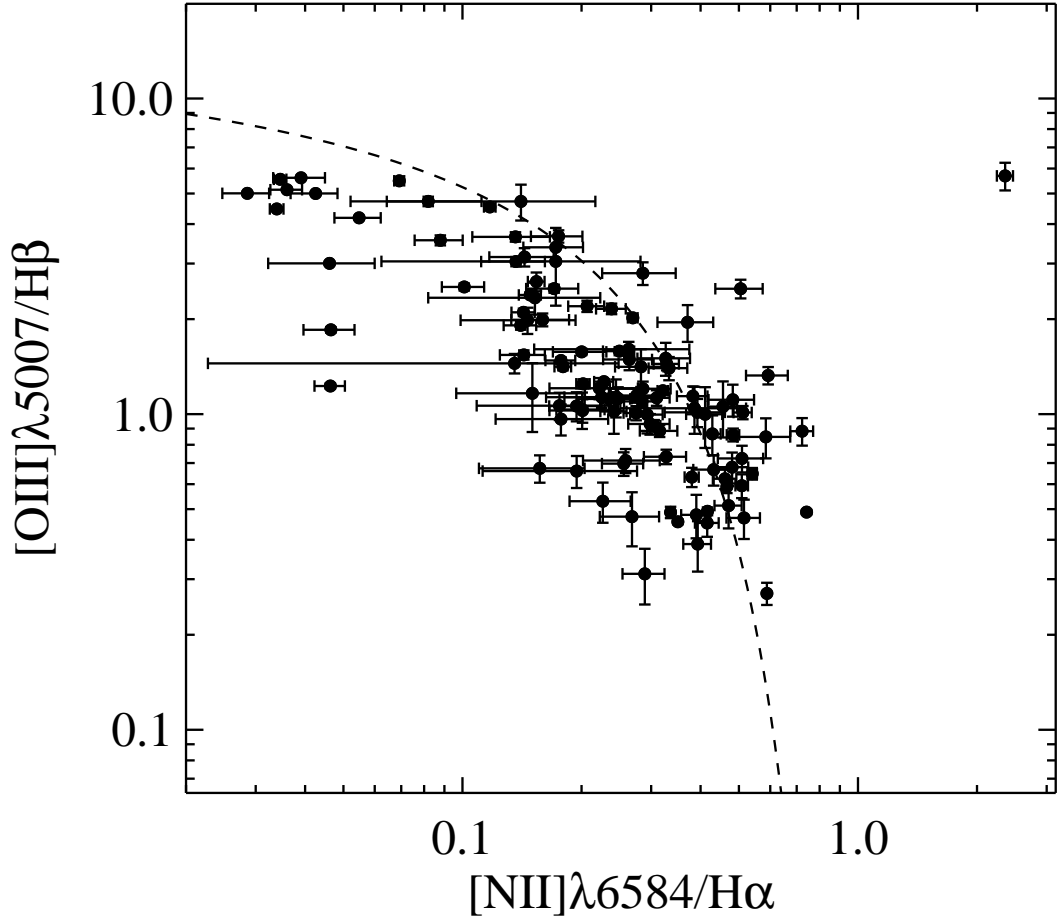}
\caption{BPT \citep{bal81} diagram for our sample of narrow-line 
objects below a redshift of 0.38; error bars are defined as the square 
root of the typical linewidth in 
pixels multiplied by the rms scatter per pixel measured from the continuum 
near the line.  The dashed line is the 
demarcation between star-forming galaxies and AGN defined by \citet{kau03}, with 
star-forming galaxies to the lower left and AGN to the right on this plot.}
\end{figure}

\begin{figure}
\plotone{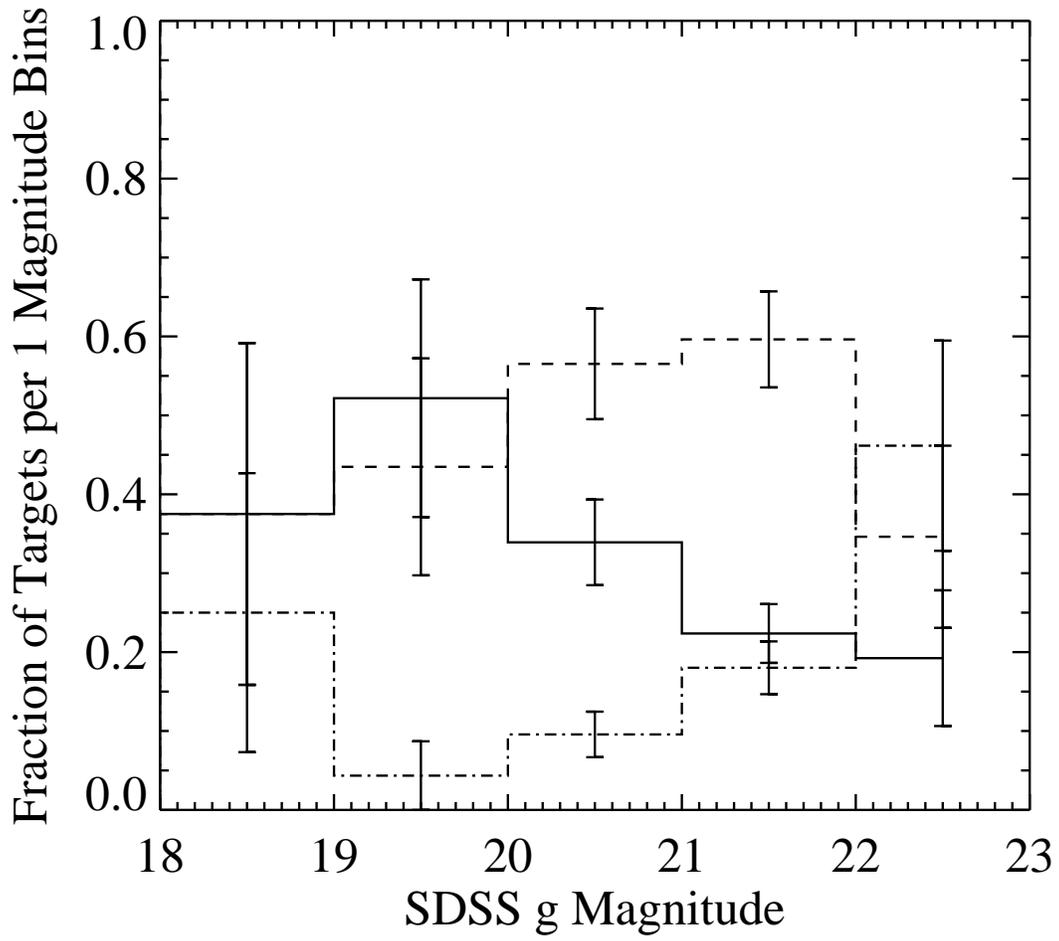}
\caption{Fraction of SDSS targets per 1 magnitude bin that are quasars (solid line), 
galaxies (dashed line), or stars/unconfirmed (dot-dashed line).}  
\end{figure}

\begin{figure}
\plotone{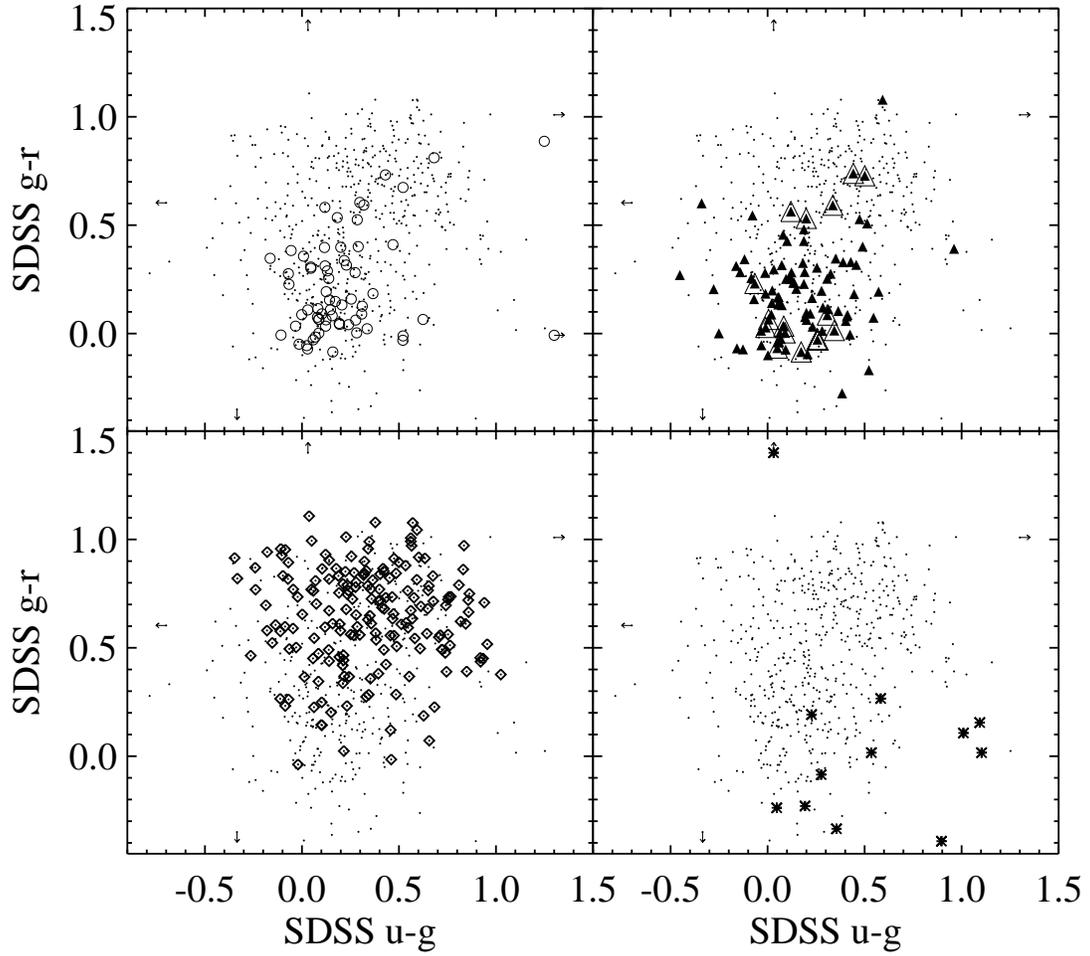}
\caption{Color-color plots of SDSS $g-r$ versus $u-g$.  (Top Left) Quasar candidates in 
this study are shown in small circles and previously confirmed SDSS quasars in large 
circles).  (Top Right)  Quasar candidates in this study are small circles and those 
confirmed are filled triangles.  Overlapping quasars between the SDSS spectroscopic follow-up and 
this study are shown as large triangles.  Improvements in photometry between DR1 and DR4 
for the previously-confirmed quasars offset points slightly 
between the top two panels.  (Bottom Left)  Quasar candidates in this study are small 
circles and those confirmed as galaxies are diamonds.  (Bottom Right)  
Quasar candidates in this study are small circles and those confirmed as stars are denoted with asterisks.}  
\end{figure}

\begin{figure}
\plotone{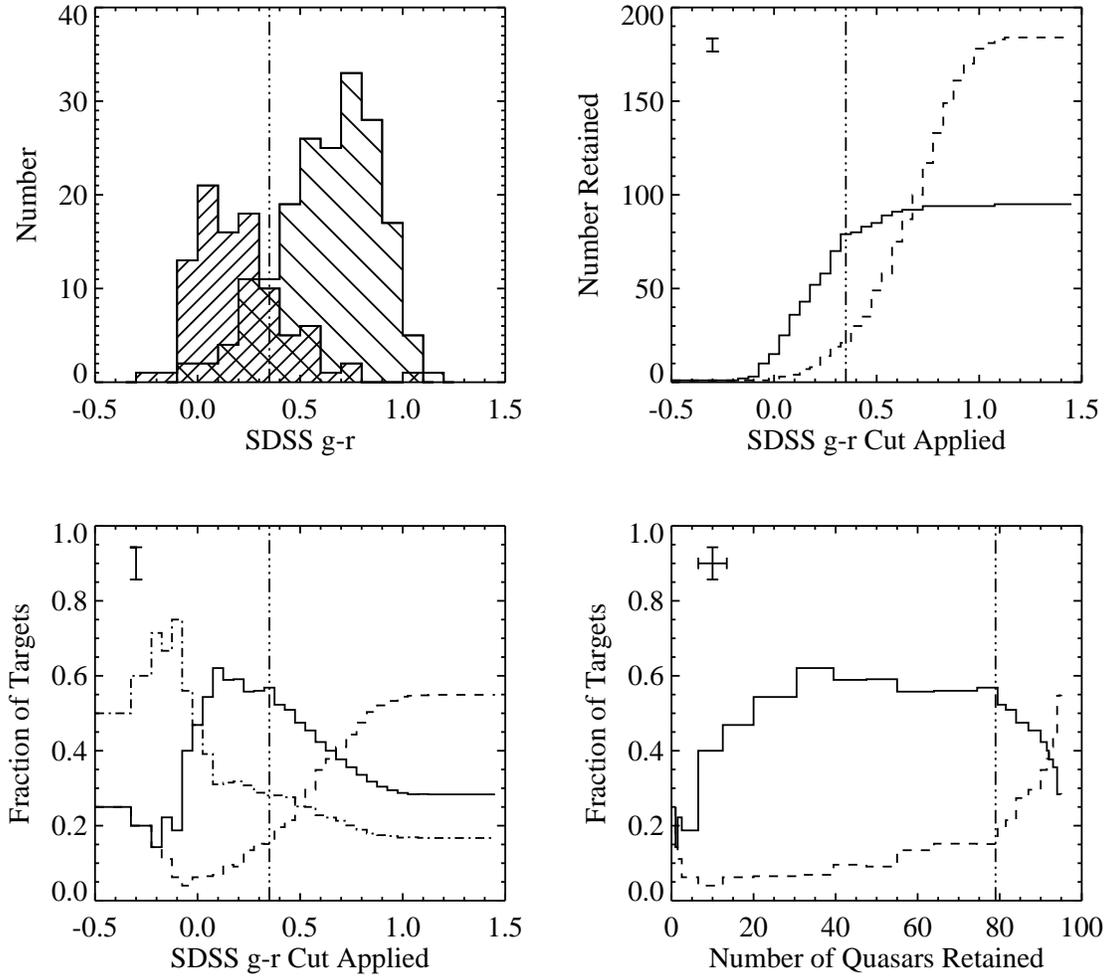}
\caption{(Upper Left) SDSS confirmed quasar and galaxy $g-r$ histograms.  Quasars are shown 
as a narrow-hatch histogram, galaxies as a wide-hatch.  The vertical line 
at $g-r = 0.35$ is the cut discussed in the text and is replicated on all panels.  
In the remaining panels quasars are shown as a solid line, galaxies as a dashed line, and 
stars/unconfirmed objects as a dot-dashed line.  The mean error bars for the quasars 
are shown in the upper left hand corner.  (Upper Right) The number of SDSS confirmed quasars and 
galaxies retained in the final sample versus the $g-r$ cut applied.  
The cut retains all objects with $g-r$ less than the given 
value.  (Lower Left) The fraction of SDSS confirmed quasars, galaxies, and stars/unconfirmed objects that 
would be retained in the sample versus the $g-r$ cut applied.  
(Lower Right)  The fraction of SDSS confirmed quasars and galaxies retained in the sample versus 
the number of quasars retained if a $g-r$ cut is applied.}  
\end{figure}

\begin{figure}
\plotone{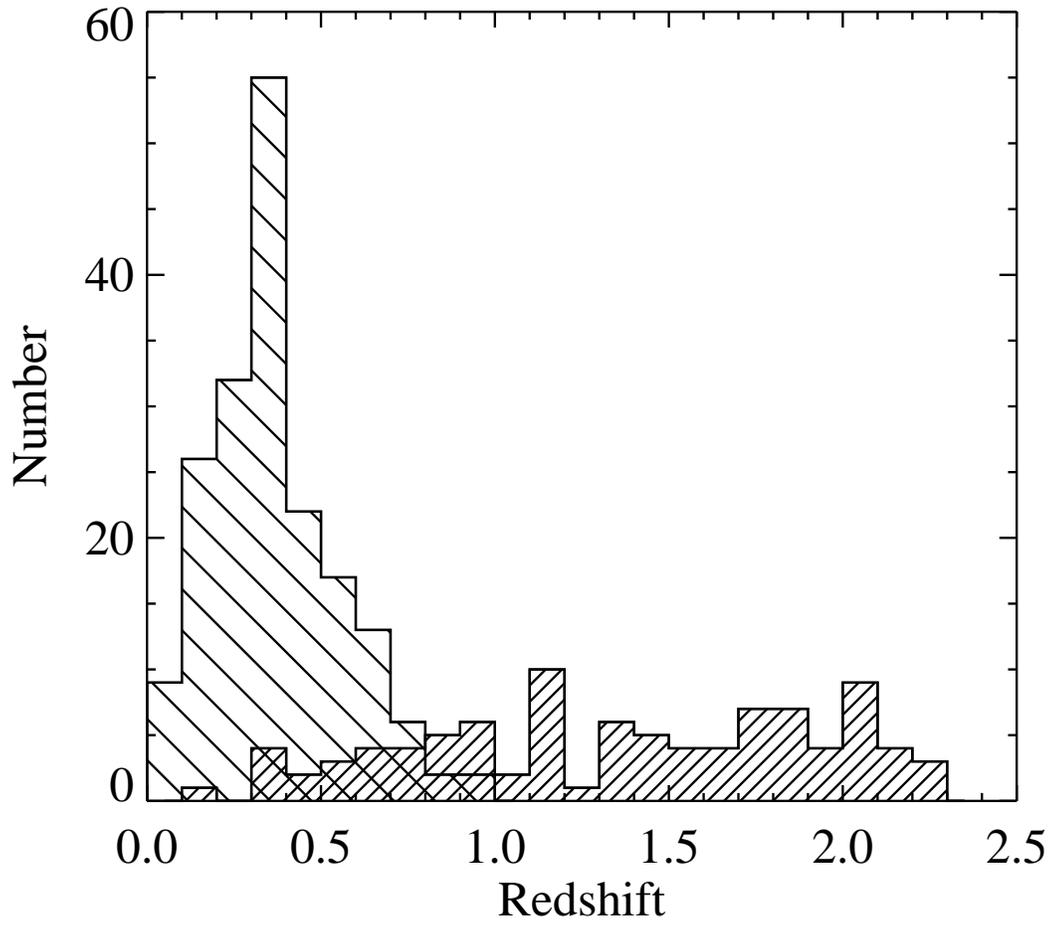}
\caption{SDSS confirmed quasar and galaxy redshift histograms.  Quasars are shown 
as a narrow-hatch histogram, galaxies as a wide-hatch.}  
\end{figure}

\begin{figure}
\plotone{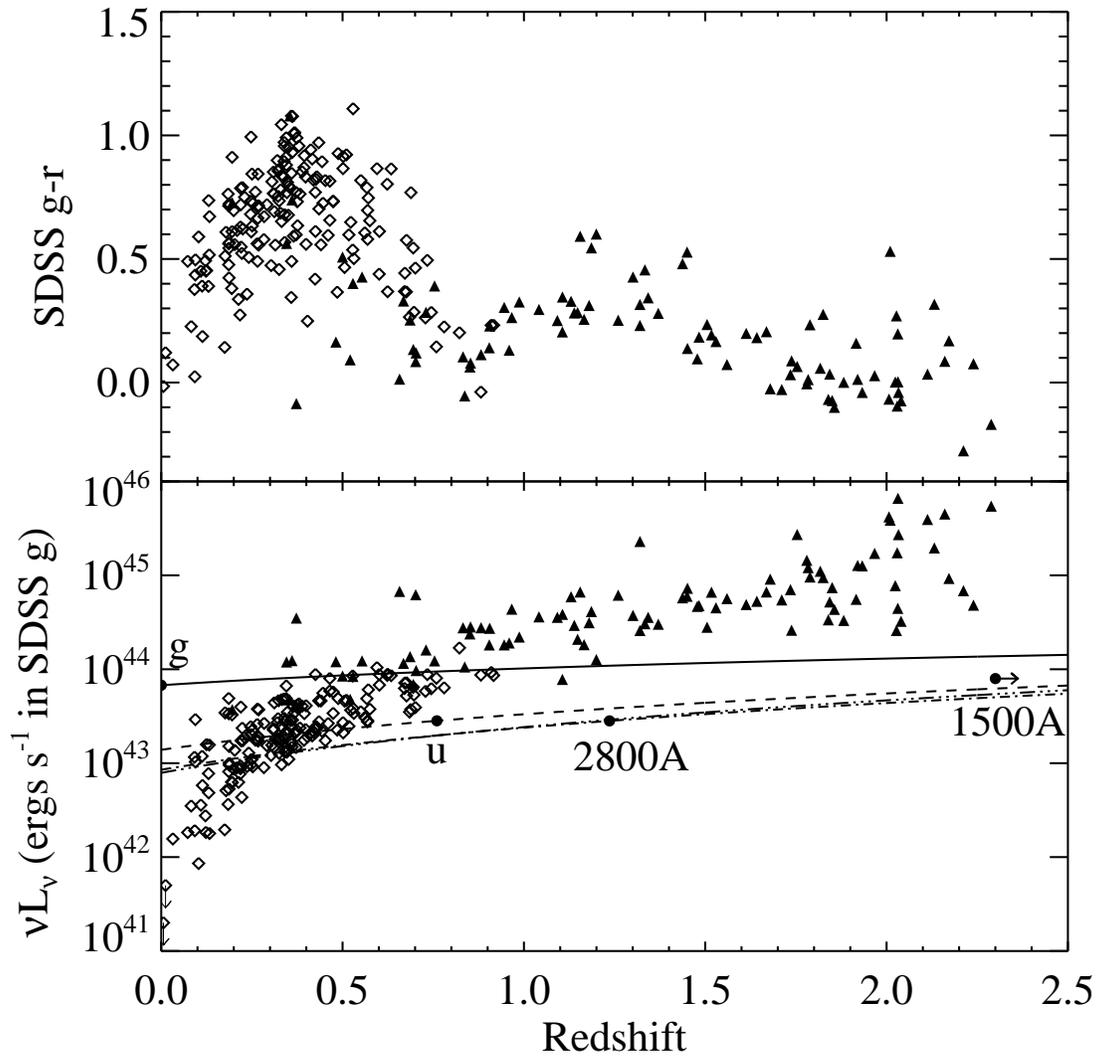}
\caption{(Top) SDSS $g-r$ versus redshift for the confirmed quasar (filled triangles) and galaxy (diamonds) 
samples in this study.  (Bottom) SDSS $g$ luminosity versus redshift for the quasar and 
galaxy samples.  The tracks show the evolution of an $L^{*}$ galaxy 
in restframe SDSS $g$ (solid), $u$ (dash), 2800~\ang\, (dot-dash), and 1500~\ang\, (dot-dot-dot-dash) from the FORS Deep Field 
(Gabasch et al. 2004).  The 2800~\ang\, and 1500~\ang\, tracks are nearly indistinguishable.  The filled circles represent 
the redshifts at which each restframe $L^{*}$ track passes through the observed g band.}  
\end{figure}

\clearpage

\appendix
\section{Discovery Spectra}
The following figures present our quasar and galaxy discovery spectra, 
given in order of Right Ascension.  Each panel is labeled with the object 
name, classification, and redshift.

\begin{figure}[h]
\figurenum{13}
\plotone{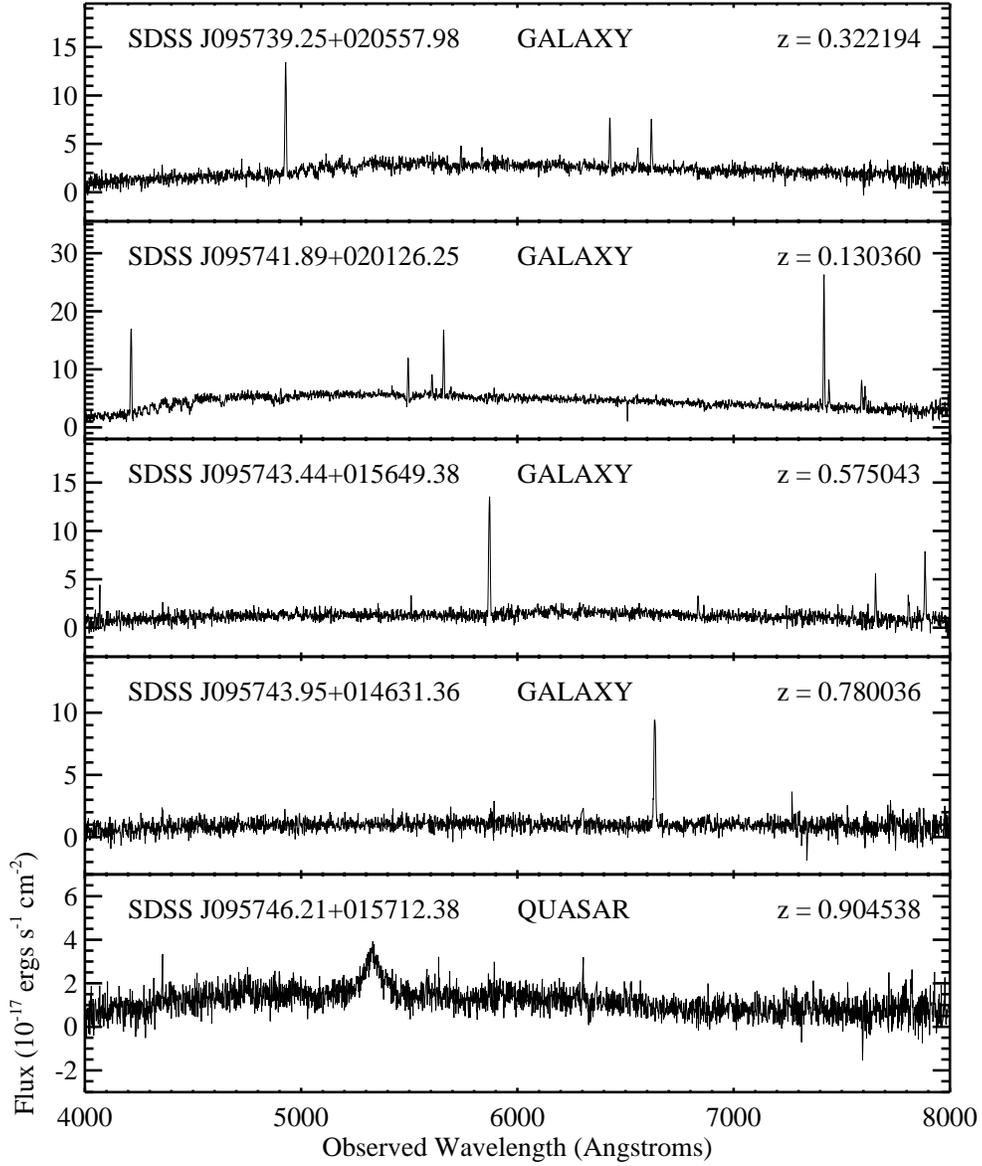}
\caption{Figures 13.1-13.56 are available in the electronic edition of 
the Journal.  The printed edition contains only a sample.}
\end{figure}

\end{document}